\documentclass[preprint]{elsarticle}

\usepackage{hyperref}

\journal{Nuclear Physics B}

\begin{document}

\begin{frontmatter}

\title{Lepton-Charge and Forward-Backward Asymmetries in Drell-Yan Processes for Precision Electroweak Measurements and New Physics Searches}

\author{J. Fiaschi\corref{mycorrespondingauthor}}
\cortext[mycorrespondingauthor]{Corresponding author}
\ead{fiaschi@liverpool.ac.uk}
\address{Department of Mathematical Sciences, University of Liverpool, Liverpool L69 3BX, United Kingdom}

\author{F. Giuli}
\ead{francesco.giuli@roma2.infn.it,francesco.giuli@cern.ch}
\address{CERN, CH 1211 Geneva}
\address{University of Rome Tor Vergata and INFN, Sezione di Roma 2, Via della Ricerca Scientifica 1, 00133 Roma}

\author{F. Hautmann}
\ead{hautmann@thphys.ox.ac.uk}
\address{Elementaire Deeltjes Fysica, Universiteit Antwerpen, B 2020 Antwerpen}
\address{Theoretical Physics Department, University of Oxford, Oxford OX1 3PU}

\author{S. Moretti}
\ead{s.moretti@soton.ac.uk}
\address{School of Physics and Astronomy, University of Southampton, Highfield, Southampton SO17 1BJ, UK}

\begin{abstract}
Precision determinations of Standard Model (SM) Electro-Weak (EW) parameters at the Large Hadron Collider (LHC) are dominated by uncertainties due to Parton Distribution Functions (PDFs).
Reweighting and profiling techniques are routinely employed to treat this. 
We explore approaches based on combining measurements of charged current and neutral current Drell-Yan (DY) asymmetries to improve PDF uncertainties. 
We present the results of a numerical analysis performed with the open-source platform {\tt{xFitter}}. 
PDF uncertainties are examined for lepton-charge and forward-backward asymmetries in regions of transverse and invariant masses near the vector-boson peak, based on LHC Run III and HL-LHC luminosity scenarios. We discuss the complementarity of the asymmetries in reducing PDF uncertainties in observables relevant to both SM and Beyond the SM (BSM) physics.
\end{abstract}


\end{frontmatter}

\section{Introduction}

The LHC, having completed its Runs I and II, has recently provided determinations of the weak mixing angle $\theta_W$~\cite{Sirunyan:2018swq,ATLAS:2018gqq,ATLAS:2018qvs} and $W$ boson mass $m_W$~\cite{Aaboud:2017svj,ATLAS:2018qzr} which are competitive in accuracy with previous determinations from lepton colliders~\cite{ALEPH:2005ab} and Tevatron~\cite{Aaltonen:2018dxj,Aaltonen:2013iut}.
The quest for ever increasing accuracy in precision measurements of EW parameters of the SM will continue to be at the center of the physics programme in forthcoming LHC runs and at the High-Luminosity LHC (HL-LHC)~\cite{Azzi:2019yne}, including the proposed Large Hadron-electron Collider (LHeC) option~\cite{Agostini:2020fmq}. 

The dominant uncertainties on current determinations of $m_W$ and $\theta_W$ at the LHC come from non-perturbative contributions due to the hadron structure in the initial state~\cite{Kovarik:2019xvh,Angeles-Martinez:2015sea}.
In the framework of QCD collinear hard-scattering factorisation for inclusive $pp$ production processes~\cite{Collins:1989gx}, such contributions are embodied in quark and gluon PDFs.
The successful completion of the precision EW physics programme at the LHC and HL-LHC thus calls for careful scrutiny of the PDFs, including understanding their correlations with the measured EW parameters, the associated uncertainties as well as looking for methods to improve the latter. 

Given this source of uncertainties, current ATLAS and CMS analyses carry out EW measurements so that PDFs are constrained in-situ by profiling and reweighting techniques.
An example is provided by the determinations of $\theta_W$ in~\cite{Sirunyan:2018swq,ATLAS:2018gqq} through DY measurements.
The CMS analysis of Ref.~\cite{Sirunyan:2018swq} uses the reweighting technique to constrain PDF uncertainties while profiling of PDF error eigenvectors is used as a cross-check. 
In the ATLAS note~\cite{ATLAS:2018gqq} the PDF uncertainties are instead included in the likelihood fit and thus constrained.

Alternative strategies focus on looking for new measurements, capable of providing high sensitivity to PDFs with low theoretical and experimental systematics while controlling correlations. An example is provided by the analysis~\cite{Abdolmaleki:2019qmq} of the neutral current DY Forward-Backward Asymmetry $A_{FB}$ in the {\tt{xFitter}} framework~\cite{Alekhin:2014irh}.
In the $Z$ boson resonance region the $A_{FB}$ observable is exploited to measure the weak mixing angle $\theta_W$ and it is sensitive to the charge-weighted linear combination $(2/3) u_V + (1/3) d_V$ of up-quark and down-quark PDFs~\cite{Abdolmaleki:2019qmq,Abdolmaleki:2019ubu} through the $Z / \gamma$ interference away from the resonance. It can thus be used to constrain quark PDFs~\cite{Accomando:2018nig,Accomando:2017scx,Fiaschi:2018buk} while additional observables are needed to achieve flavour decomposition. 

The purpose of this work is to extend the investigation in~\cite{Abdolmaleki:2019qmq} by using the charged current lepton-charge asymmetry $A_W$ in combination with $A_{FB}$ and study the implications of the combined analysis on PDF uncertainties.
We will exploit the sensitivity of the lepton-charge asymmetry $A_W$ to the difference $u_V - d_V$ to probe linearly independent combinations of up-quark and down-quark PDFs from both $A_{FB}$ and $A_W$.

We will consider measurements of lepton-charge asymmetry at Runs I and II~\cite{Aad:2019rou,Sirunyan:2020oum} as well as future measurements at Run III and HL-LHC luminosities. 

Studies of the impact of $W$ and $Z$ production data on PDFs have recently appeared~\cite{Fu:2020mxl,Deng:2020sol} based on the {\sc ePump} package~\cite{Hou:2019gfw,Schmidt:2018hvu,Willis:2018yln}. 

The paper is organised as follows.
In Sec.~II we describe the calculational framework for DY asymmetries in {\tt{xFitter}} and validate it by comparing Next-to-Leading-Order (NLO) results with Run I experimental data.
In Sec.~III we consider Run III and HL-LHC luminosities and perform a PDF profiling calculation.
Using {\tt{xFitter}}, we analyze the separate and combined impact of high-statistics $A_{FB}$ and $A_W$ asymmetry measurements in the mass region near the vector boson pole on PDF uncertainties.
In Sec.~IV we illustrate the implications of this analysis for various DY observables, discussing examples both in the region of SM vector boson masses and in the multi-TeV region relevant for new physics BSM searches. We give our conclusions in Sec.~V.

\section{xFitter calculational framework and comparison with Run I data}
\label{AW_8TeV_peak}

We here recall the main elements of the {\tt{xFitter}}~\cite{Alekhin:2014irh} calculational framework applied to the DY lepton-charge asymmetry $A_W$, defined as

\begin{equation}
 A_W = \frac{d\sigma/d|\eta_\ell|(W^+ \rightarrow \ell^+ \nu) - d\sigma/d|\eta_\ell|(W^- \rightarrow \ell^- \bar{\nu})}{d\sigma/d|\eta_\ell|(W^+ \rightarrow \ell^+ \nu) + d\sigma/d|\eta_\ell|(W^- \rightarrow \ell^- \bar{\nu})},
\end{equation}

\noindent
and to the reconstructed DY neutral forward-backward asymmetry $A_{FB}$, defined as

\begin{equation}
 A_{FB} = \frac{d\sigma/dM_{\ell\ell}(\cos\theta^* > 0) - d\sigma/dM_{\ell\ell}(\cos\theta^* < 0)}{d\sigma/dM_{\ell\ell}(\cos\theta^* > 0) + d\sigma/dM_{\ell\ell}(\cos\theta^* < 0)},
\end{equation}

\noindent
where $\eta_\ell$ is the pseudorapidity of the charged lepton defined in the laboratory frame, $M_{\ell\ell}$ is the di-lepton system invariant mass and $\theta^*$ is the angle between the outgoing lepton and the incoming quark defined in the Collins–Soper frame~\cite{Collins:1977iv}.

We validate the implementation of $A_W$ by performing fits to ATLAS experimental data~\cite{Aad:2019rou} at $\sqrt{s} = $ 8 TeV.
The observable has been computed at NLO in perturbative QCD, using the {\tt{MadGraph5\_aMC@NLO}}~\cite{Alwall:2014hca} program, interfaced to {\tt{APPLgrid}}~\cite{Carli:2010rw} through {\tt{aMCfast}}~\cite{Bertone:2014zva}. We obtain theoretical predictions corresponding to the analysis cuts of the ATLAS data recorded at $\sqrt{s}$ = 8 TeV from Ref.~\cite{Aad:2019rou}. The renormalisation and factorisation scales $\mu_{R}$ and $\mu_{F}$ in the NLO computations are set equal to $\mu_{R} = \mu_{F} = m_{W}$.

These computations are supplemented by $K$-factors to match theoretical predictions from an optimised version of the {\tt{DYNNLO}} generator~\cite{Catani:2009sm}, which simulates initial-state QCD corrections to Next-to-NLO (NNLO) accuracy, at LO in the EW couplings with parameters set according to the $G_{\mu}$ scheme~\cite{Hollik:1988ii}. The input parameters (the Fermi constant $G_{\mathrm{F}}$, the masses and widths of the $W$ and $Z$ bosons and the Cabibbo-Kobayashi-Maskawa (CKM) entries) are taken from Ref.~\cite{Patrignani:2016xqp}. 

\begin{figure}
\begin{center}
\includegraphics[width=0.5\textwidth]{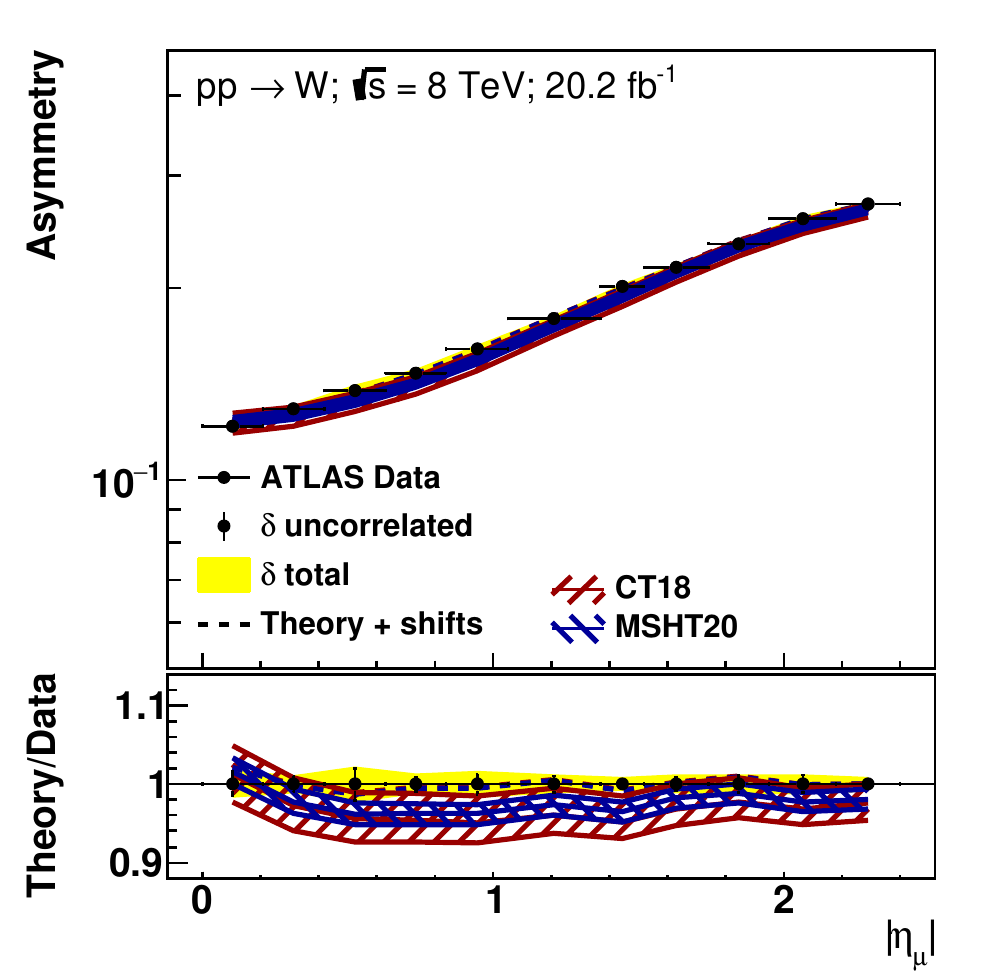}
\end{center}
\caption{Muon rapidity dependence of the lepton-charge asymmetry $A_W$, 
for CT18NNLO~\protect\cite{Hou:2019efy} and MSHT20~\protect\cite{Bailey:2020ooq} PDFs, compared with the ATLAS measurements~\protect\cite{Aad:2019rou} at $\sqrt{s} = $ 8 TeV.}
\label{fig:AW_ATLAS_8TeV}
\end{figure}

In Fig.~\ref{fig:AW_ATLAS_8TeV} we implement the calculation of $A_W$ in {\tt{xFitter}}~\cite{Alekhin:2014irh} and present the comparison of our theoretical predictions for $A_W$ (for two choices of PDFs, CT18NNLO~\cite{Hou:2019efy} and MSHT20nnlo~\cite{Bailey:2020ooq}) with the ATLAS experimental measurements~\cite{Aad:2019rou}.
In Tab.~\ref{tab:AW_fit} we extend the comparison by performing fits with several PDF sets: CT18NNLO, CT18ANNLO~\cite{Hou:2019efy}, MSHT20nnlo~\cite{Bailey:2020ooq}, NNPDF3.1nnlo~\cite{Ball:2017nwa}, PDF4LHC15nnlo~\cite{Butterworth:2015oua}, ABMP16nnlo~\cite{Alekhin:2017kpj} and HERAPDF2.0nnlo~\cite{Abramowicz:2015mha}. The results for the $\chi^{2}$ values are reported in Tab.~\ref{tab:AW_fit} for each of the PDF sets, showing a good description of data for all sets.

\begin{table}
\begin{center}
\begin{tabular}{|c|c|}
  \hline
  \textbf{PDF set} & \textbf{$\chi^2$/d.o.f.} \\
  \hline
  CT18NNLO & 10.26/11\\
  \hline
  CT18ANNLO & 11.29/11\\
  \hline
  MSHT20nnlo\_as118 & 12.18/11\\
  \hline
  NNPDF3.1\_nnlo\_as\_0118\_hessian & 14.88/11\\
  \hline
  PDF4LHC15\_nnlo\_100 & 9.53/11\\
  \hline
  ABMP16\_5\_nnlo & 18.21/11\\
  \hline
  HERAPDF20\_NNLO\_EIG & 8.92/11\\
  \hline
\end{tabular}
\caption{The $\chi^2$ values per degree of freedom from fits to $A_W$ experimental 
measurements~\protect\cite{Aad:2019rou} using {\tt{xFitter}}~\protect\cite{Alekhin:2014irh}, for 
different PDF sets. PDF uncertainties are evaluated at the 68\% Confidence Level (CL).}
\label{tab:AW_fit}
\end{center}
\end{table}

\begin{figure}
\begin{center}
\includegraphics[width=0.5\textwidth]{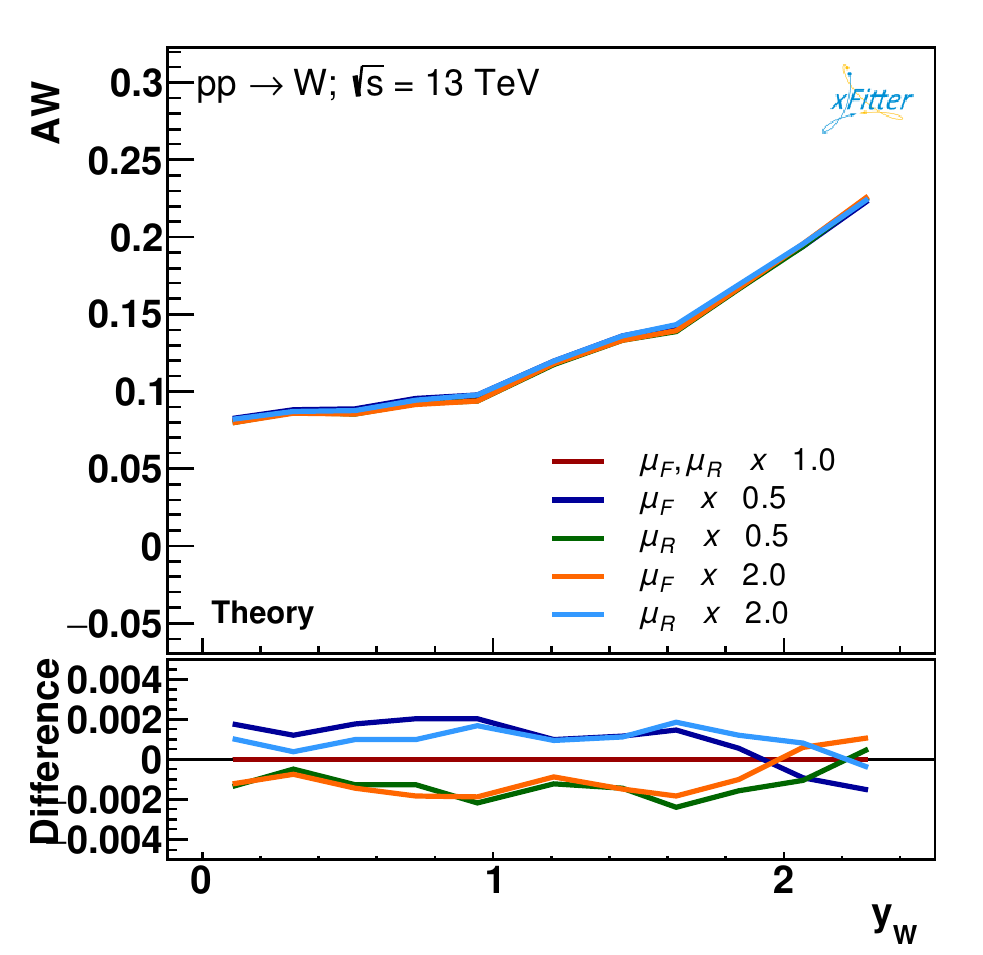}
\end{center}
\caption{$W$-boson rapidity dependence of the lepton-charge asymmetry $A_W$ for different choices of factorisation and renormalisation scales.}
\label{fig:AW_scales}
\end{figure}

We investigate the accuracy of our computation for $A_W$ by estimating its theory uncertainty through variation of the factorisation and renormalisation scales.
Fig.~\ref{fig:AW_scales} shows $A_W$ theoretical predictions versus $W$-boson rapidity 
at $\sqrt{s} = $ 13 TeV, where the factorisation and renormalisation scales are varied by a factor 2 or 0.5.
The differences with respect to the $A_W$ central value curve ($\mu_F$ and $\mu_R$ unchanged) are at the permille level, thus confirming the reliability of the calculations with respect to this systematic uncertainty.

\section{Complementarity of ${A_{FB}}$ and ${A_W}$}
\label{sec:compl}

In this section we extend the profiling analysis~\cite{Abdolmaleki:2019qmq} of $A_{FB}$ by using $A_W$ in combination with $A_{FB}$, and study the implications of the combined analysis on PDF uncertainties.
We generate $A_{FB}$ and $A_W$ pseudodata at $\sqrt{s} = $ 13 TeV with integrated luminosities of 
300 and 3000 fb$^{-1}$, corresponding respectively to the designed values at the end of the LHC Run III and at the end of the HL-LHC.

We employ pseudodata for $A_{FB}$ in the mass region around the $Z$ peak, and pseudodata for $A_W$ 
in the mass region around the $W$ Jacobian peak. We compute $A_W$ as described in Sec.~\ref{AW_8TeV_peak}.
For $A_{FB}$ we carry out an analogous calculation as described in Ref.~\cite{Abdolmaleki:2019qmq}.
We adopt the analysis cuts of the ATLAS $A_{FB}$ measurements~\cite{Aaboud:2017ffb} at $\sqrt{s}$ = 8 TeV, in the invariant mass interval 45 GeV $\leq M_{\ell\ell} \leq $ 200 GeV using 61 bins of 2.5 GeV width, and set the $\mu_{R}$ and $\mu_{F}$ scales to the invariant mass of the dilepton pair in the final state, namely, $\mu_{R} = \mu_{F} = m_{\ell\ell}$.
We project the statistical error on these observables for the luminosity scenarios considered by combining the statistics of electron and muon channels.

Following the method of Refs.~\cite{Abdolmaleki:2019qmq,Amoroso:2020fjw}, we apply the technique~\cite{Paukkunen:2014zia,Camarda:2015zba} to evaluate PDF uncertainties. 
The results will be shown at the chosen representative energy scale $Q^2 = M_Z^2 =$ 8317 GeV$^2$. We have checked that the qualitative behavior of the profiled distributions does not change when varying the $Q^2$ values.

\subsection{Eigenvector rotation}

We first want to identify the PDFs and their combinations which are most sensitive to the $A_W$ observable, by performing an eigenvector rotation exercise analogous to that done for the case of $A_{FB}$ in Ref.~\cite{Abdolmaleki:2019qmq}.
Using this technique~\cite{Pumplin:2009nm}, we rotate and sort the eigenvectors of the CT18NNLO PDF set according to the $A_W$ pseudodata's impact on them, and plot the contribution of the first 4 eigenvectors to the error bands of different PDFs and their combinations.

\begin{figure}
\begin{center}
\includegraphics[width=0.3\textwidth]{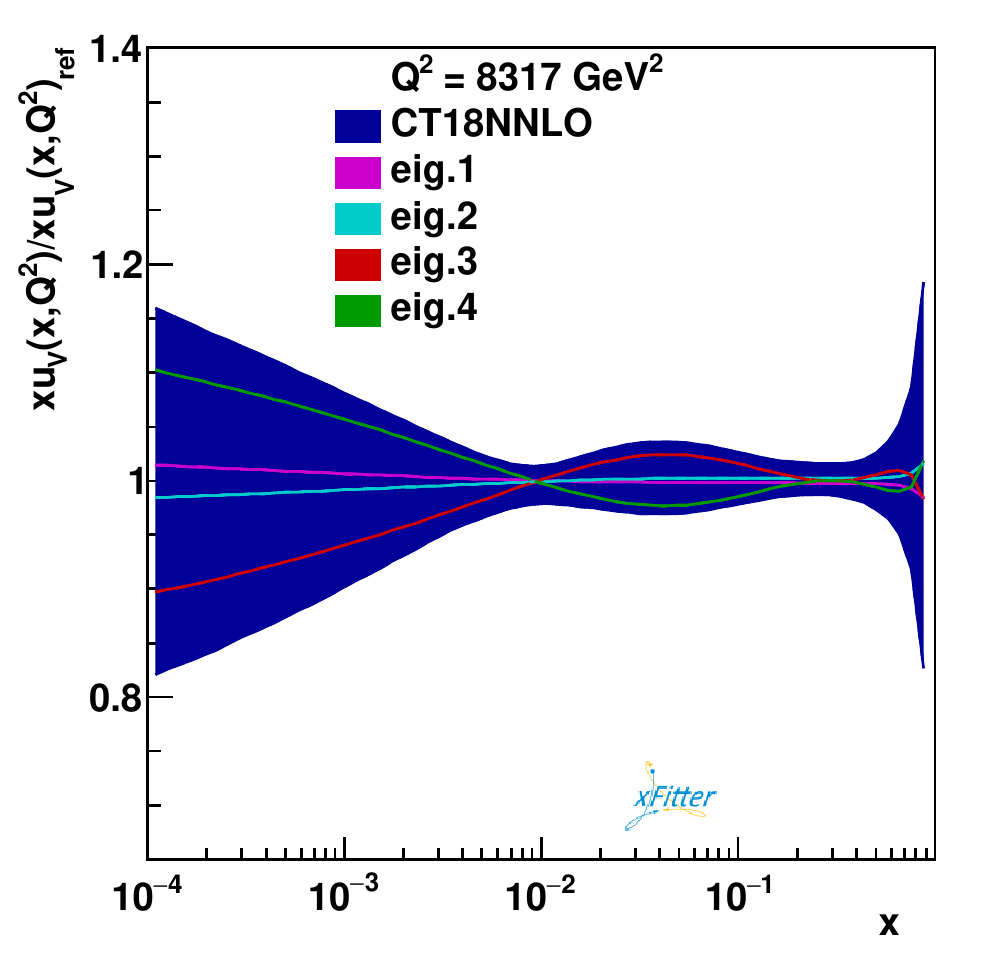}
\includegraphics[width=0.3\textwidth]{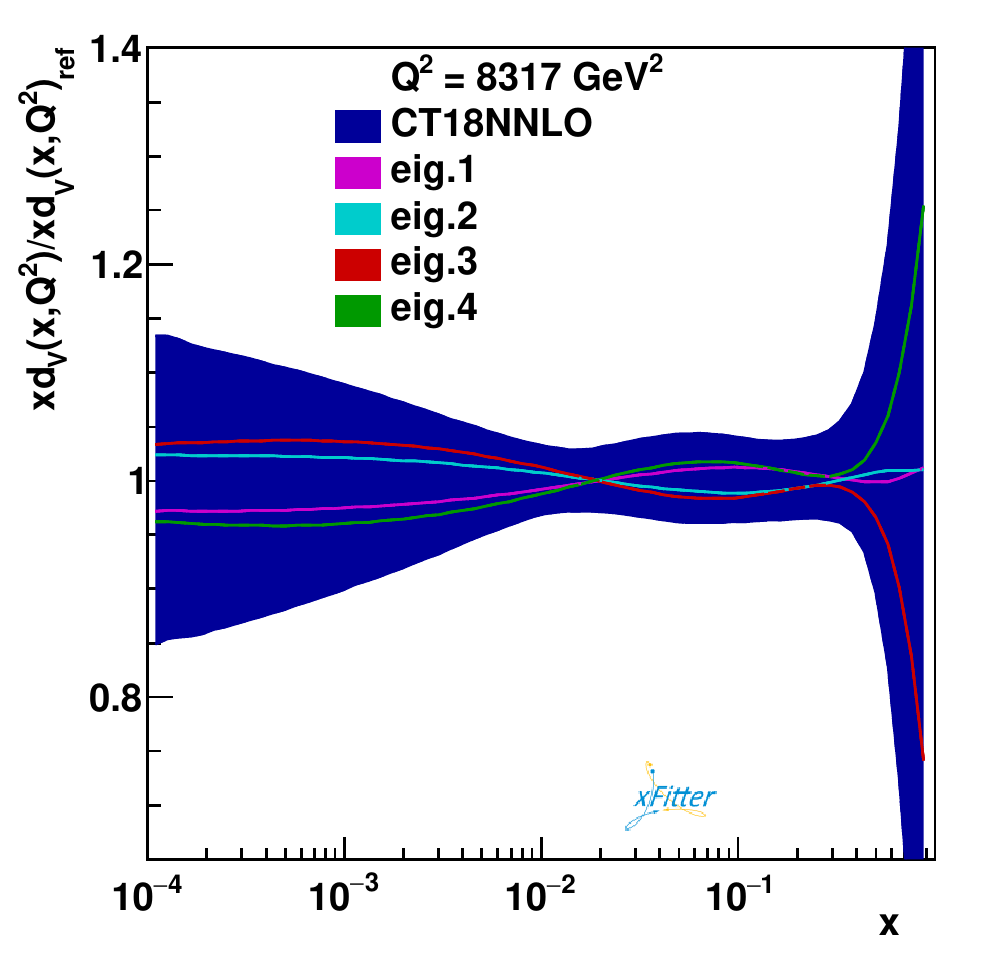}
\includegraphics[width=0.3\textwidth]{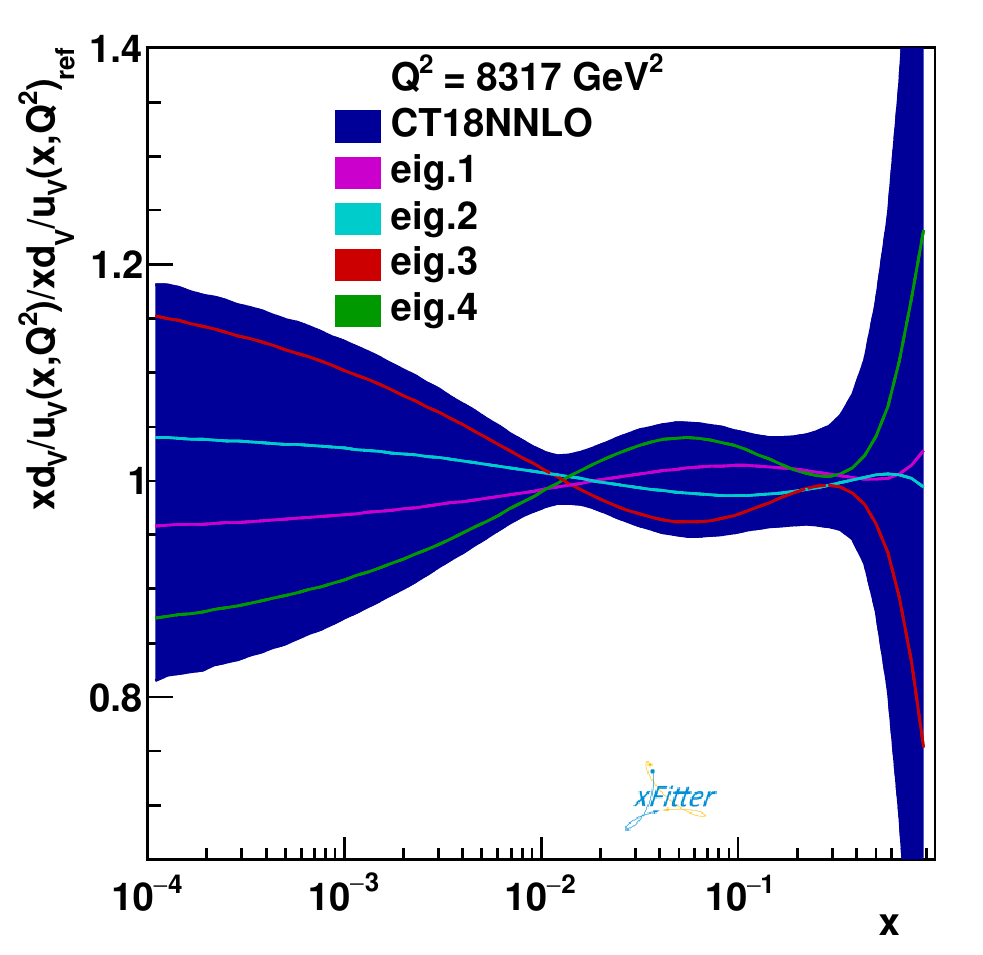}\\
\includegraphics[width=0.3\textwidth]{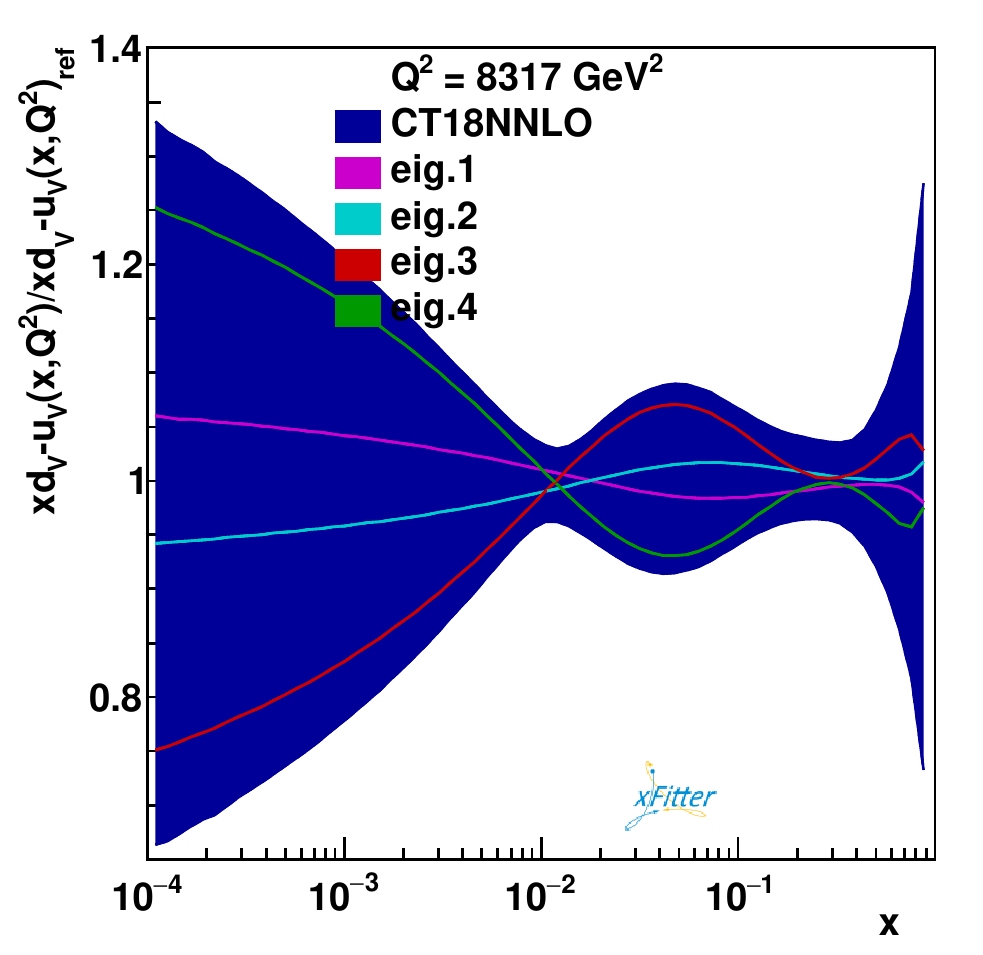}
\includegraphics[width=0.3\textwidth]{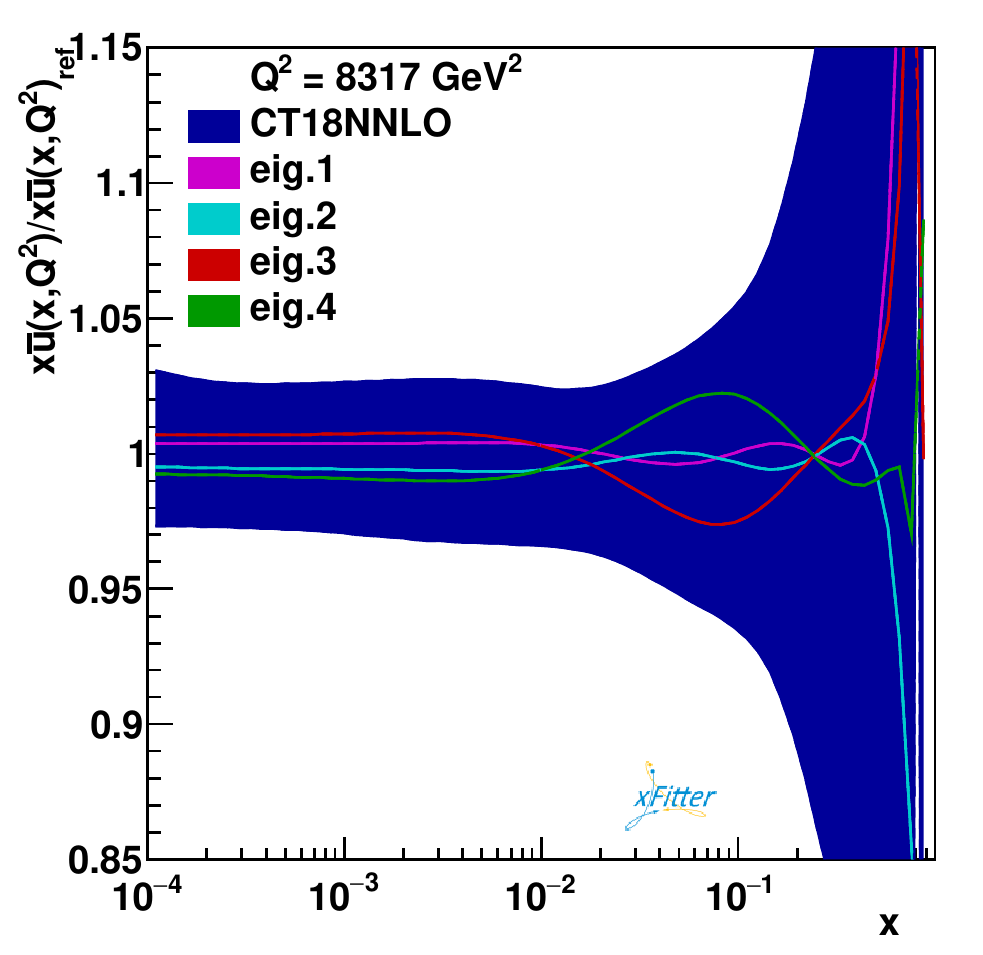}
\includegraphics[width=0.3\textwidth]{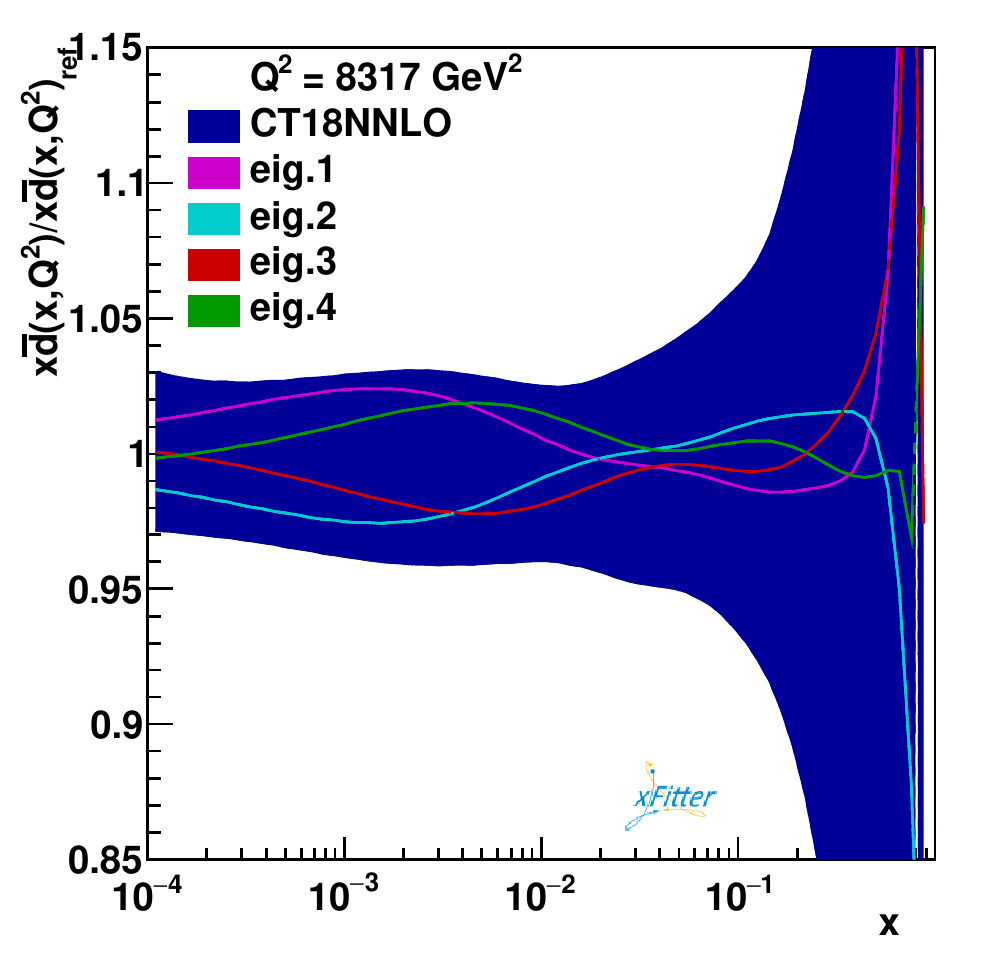}
\end{center}
\caption{Contribution of rotated eigenvectors to the various PDFs.}
\label{fig:AW_rot}
\end{figure}

Fig.~\ref{fig:AW_rot} shows that the third and fourth eigenvectors provide the largest contribution to quark PDFs, while the first two eigenvectors have a comparable weight to the third and the fourth in the anti-quark PDFs.
The saturation given by the first four eigenvectors on the error bars of the PDF composition $d_V - u_V$ confirms that the $A_W$ observable is most sensitive to this combination, and similar conclusion can be drawn on the PDF ratio $d_V / u_V$.
In the following, we will then show the results of the profiling on these PDF distributions as well.

\subsection{Profiling with $A_{FB}$ and $A_{W}$ separately}
\label{sec:data_on_peak}

\begin{figure}
\begin{center}
\includegraphics[width=0.24\textwidth]{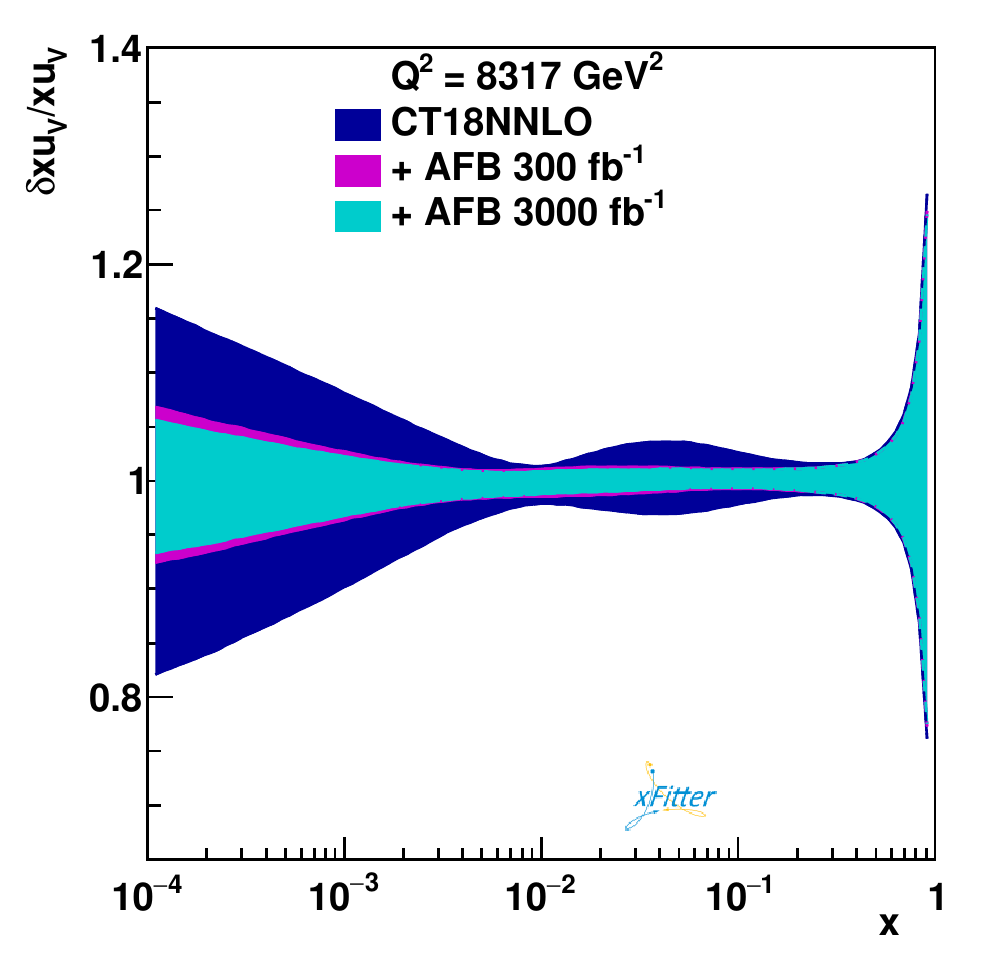}
\includegraphics[width=0.24\textwidth]{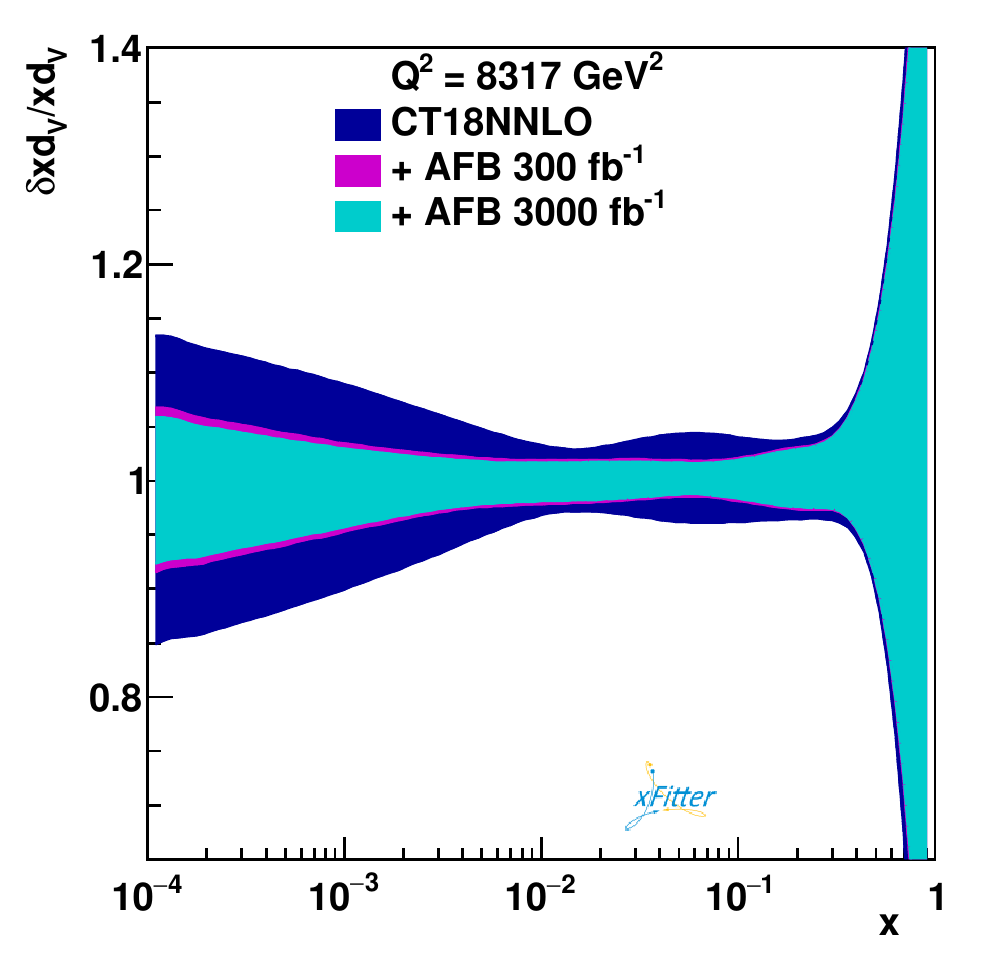}
\includegraphics[width=0.24\textwidth]{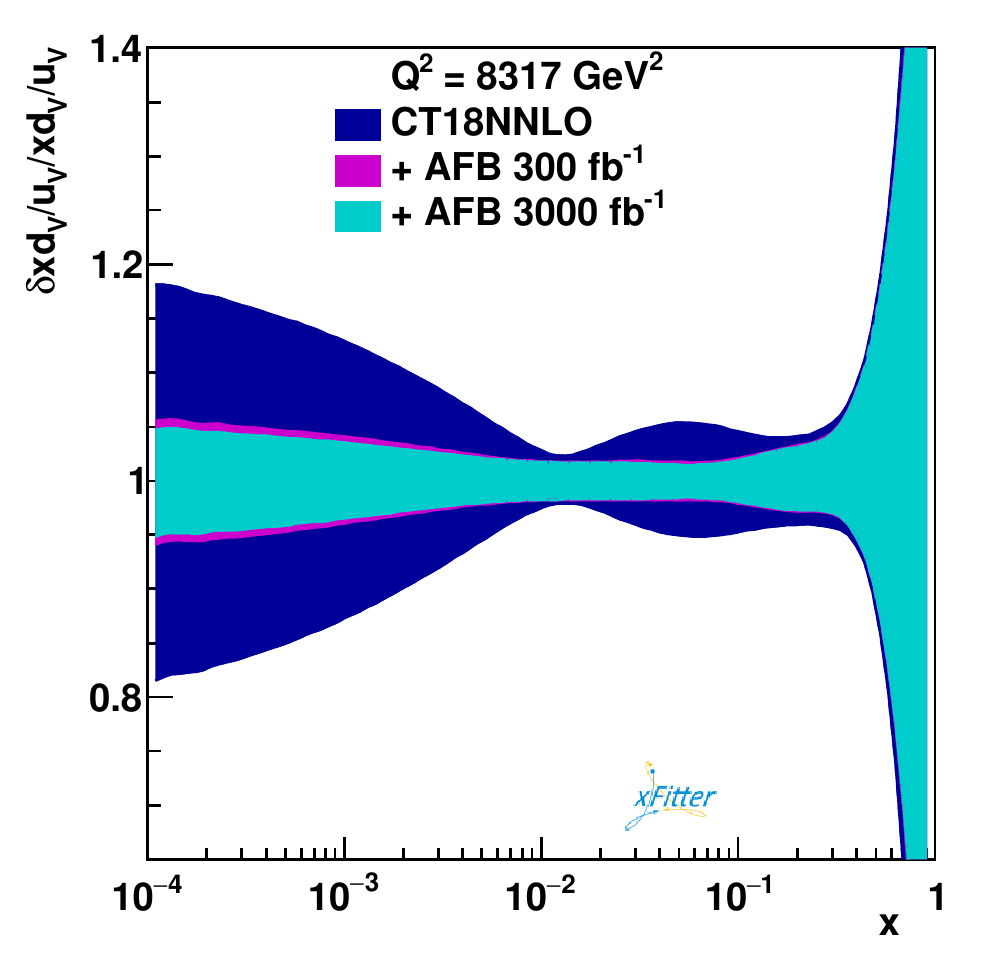}
\includegraphics[width=0.24\textwidth]{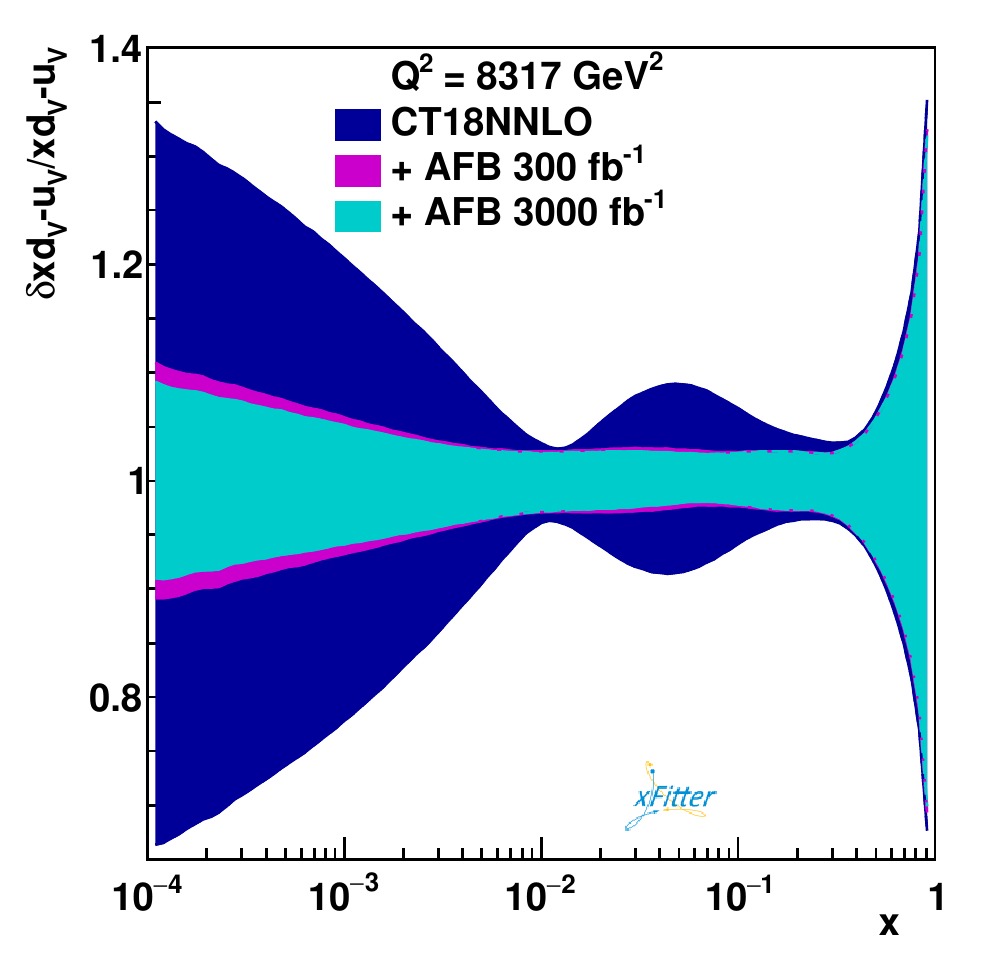}\\
\includegraphics[width=0.24\textwidth]{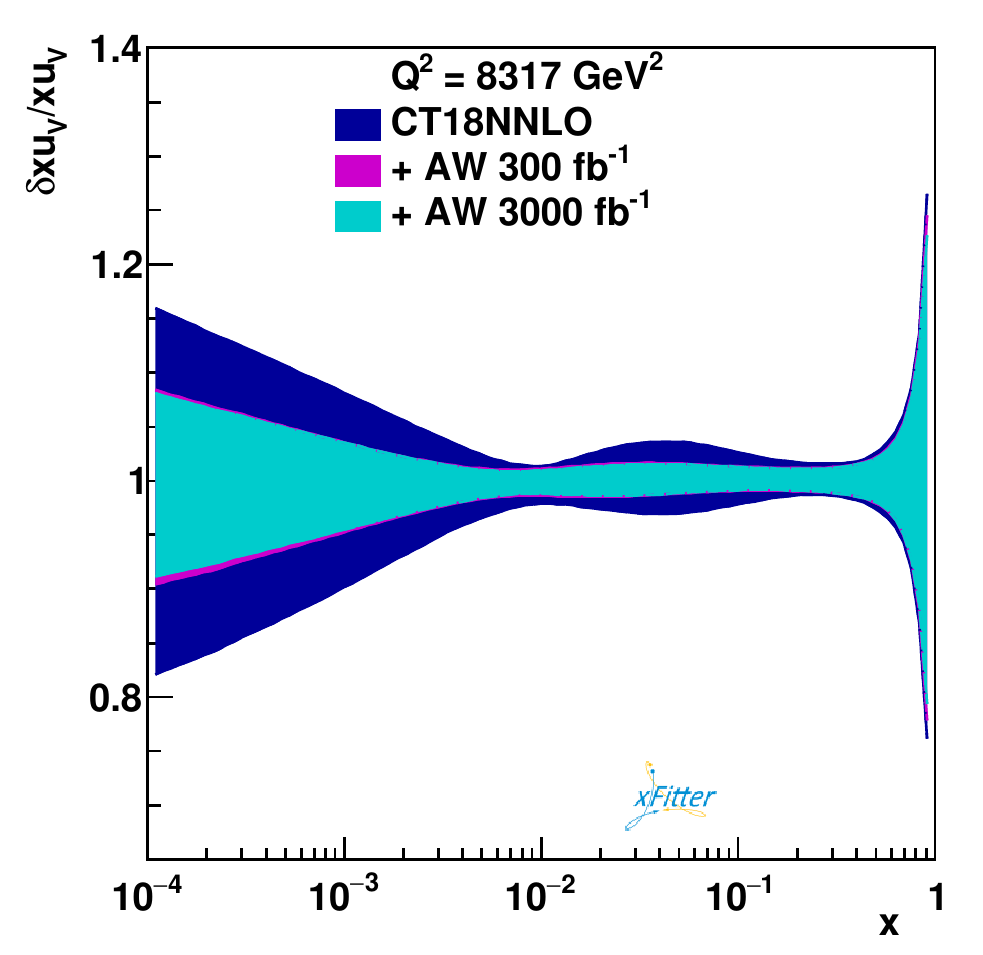}
\includegraphics[width=0.24\textwidth]{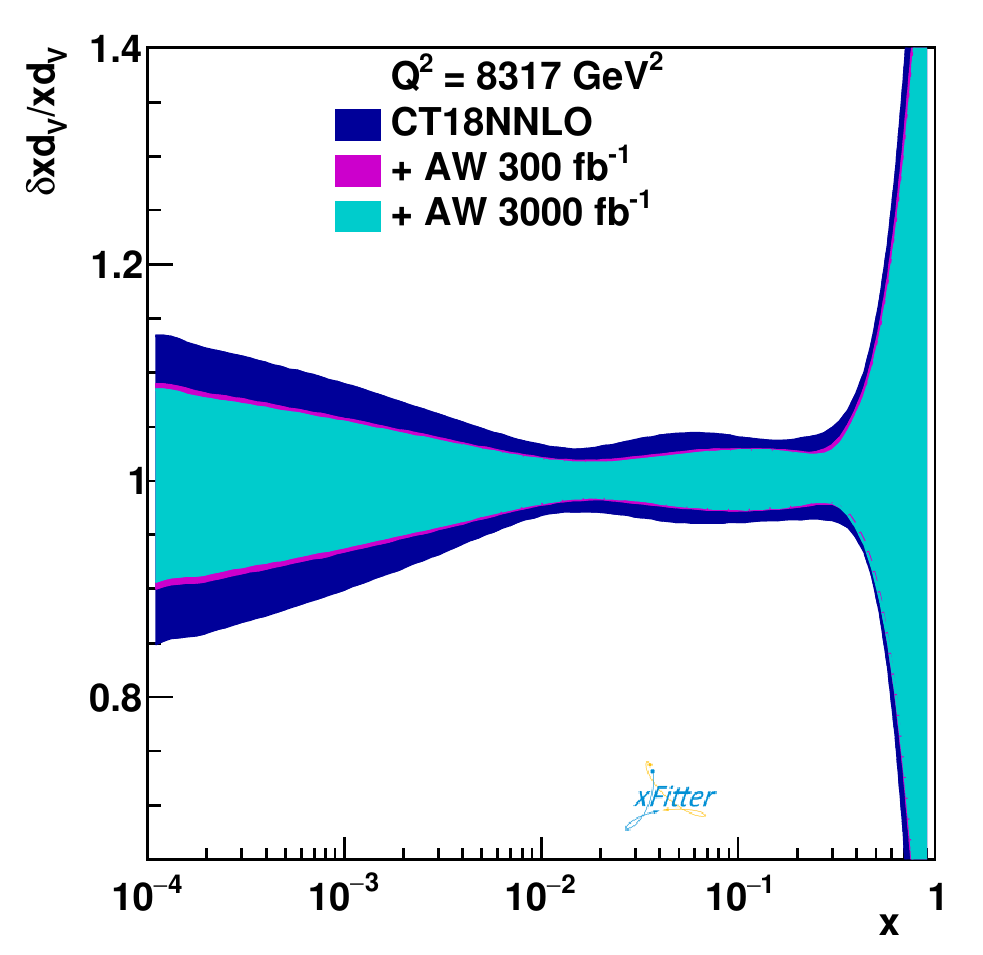}
\includegraphics[width=0.24\textwidth]{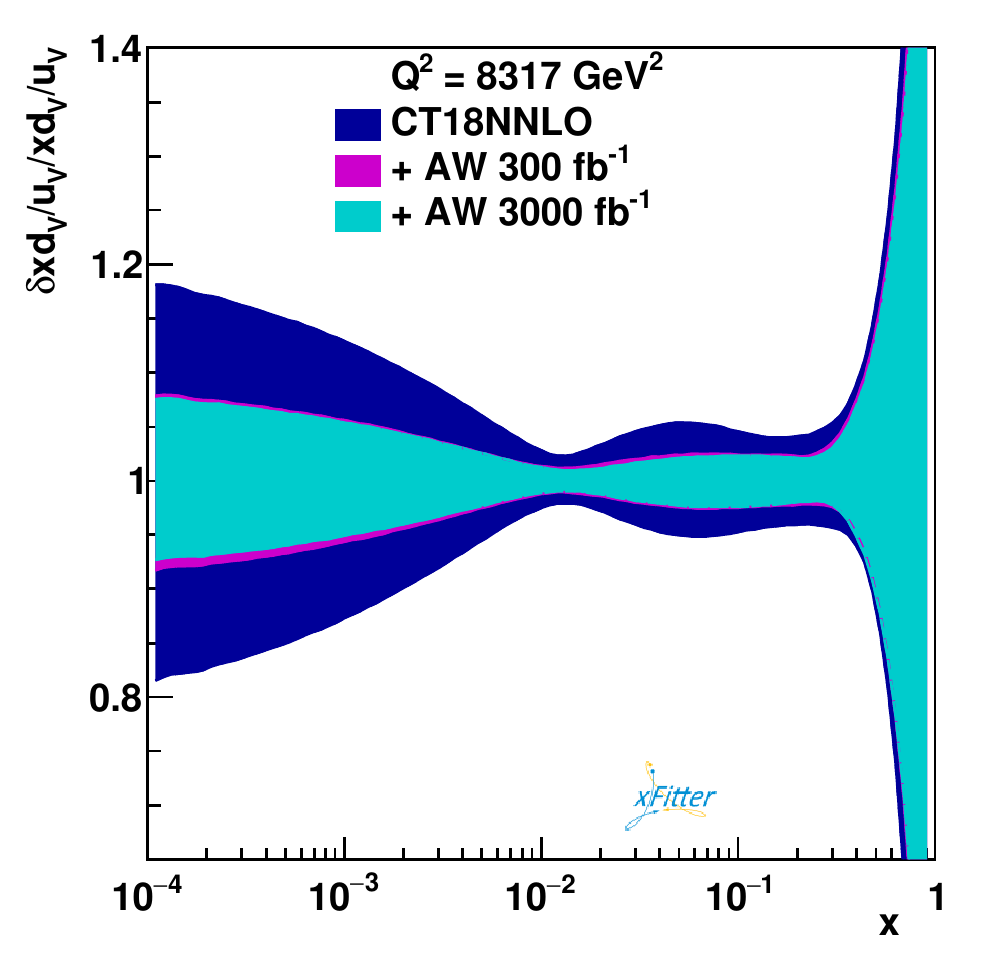}
\includegraphics[width=0.24\textwidth]{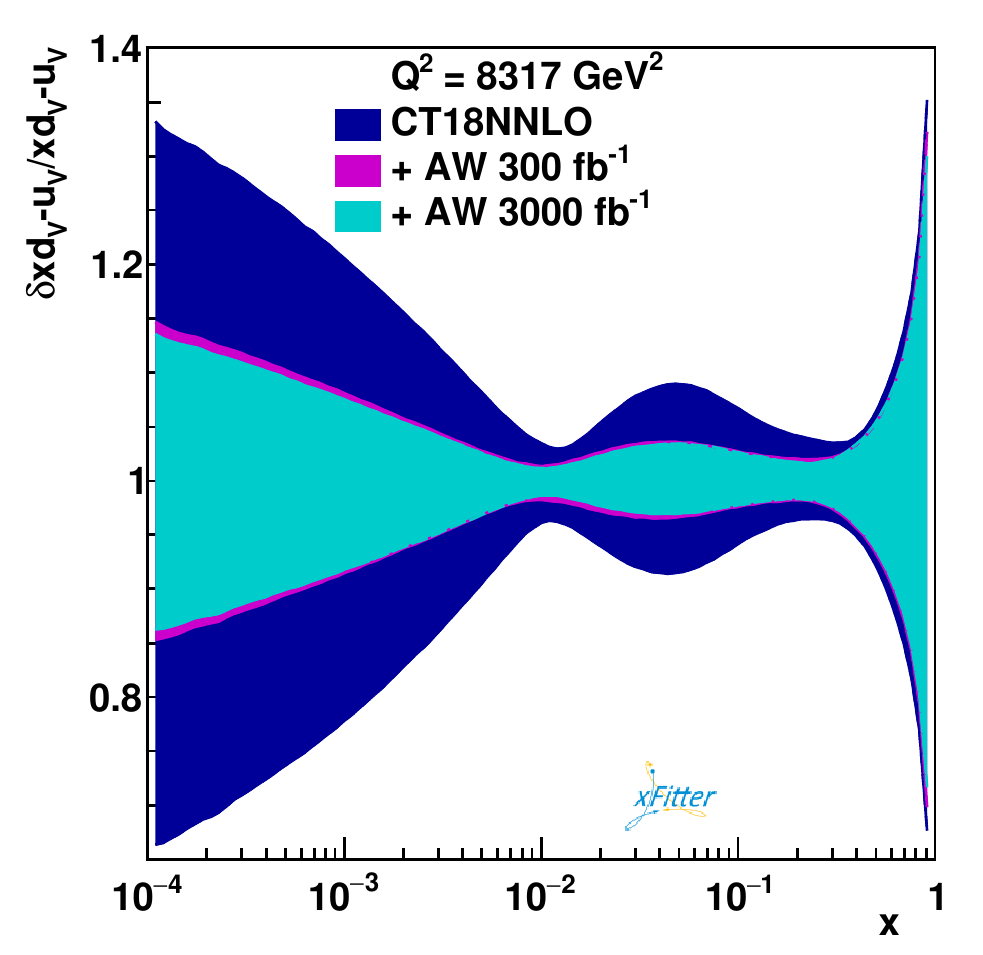}
\end{center}
\caption{Original CT18NNLO~\cite{Hou:2019efy} (blue) and profiled distributions using $A_{FB}$ (top) and $A_W$ (bottom) pseudodata corresponding to integrated luminosities of 300 fb$^{-1}$ (pink) and 3000 fb$^{-1}$ (cyan). Results are shown for valence quark distributions at $Q^2 = M_Z^2 =$ 8317 GeV$^2$.}
\label{fig:AFB_AW_valence}
\end{figure}

We show here the reduction of PDF error bands when $A_{FB}$ and $A_W$ pseudodata separately are used in the profiling.
The constraints placed by $A_{FB}$ and $A_W$ on the valence quark PDFs are shown in Fig.~\ref{fig:AFB_AW_valence}. We note that the reduction of the error bands given by the two observables are comparable, with $A_{FB}$ providing slightly stronger constraints.

\begin{figure}
\begin{center}
\includegraphics[width=0.3\textwidth]{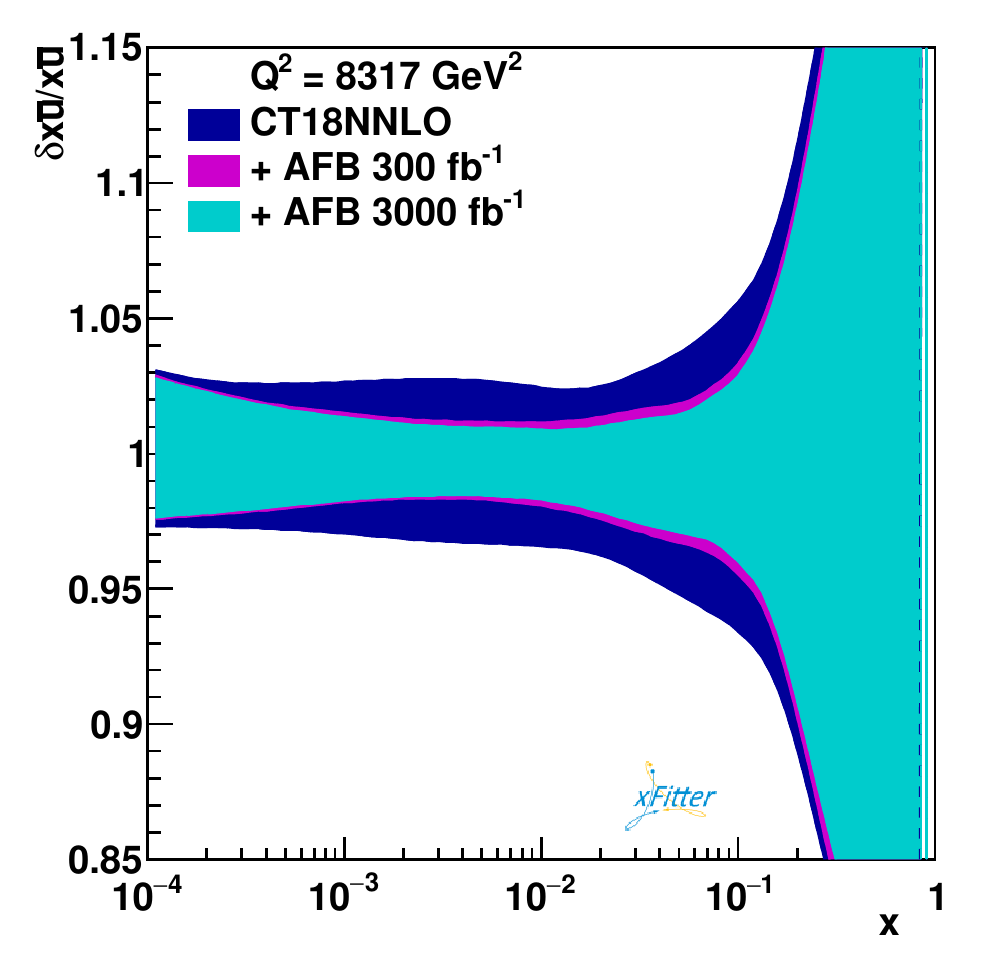}
\includegraphics[width=0.3\textwidth]{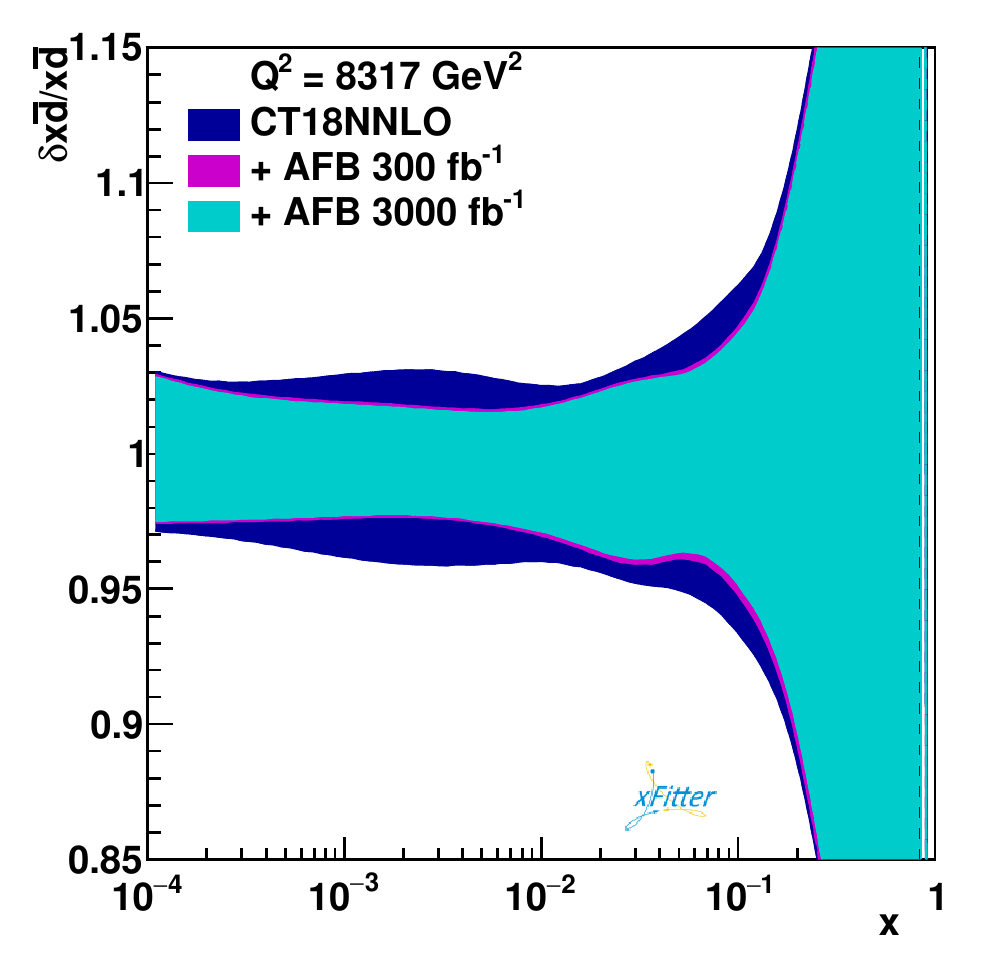}\\
\includegraphics[width=0.3\textwidth]{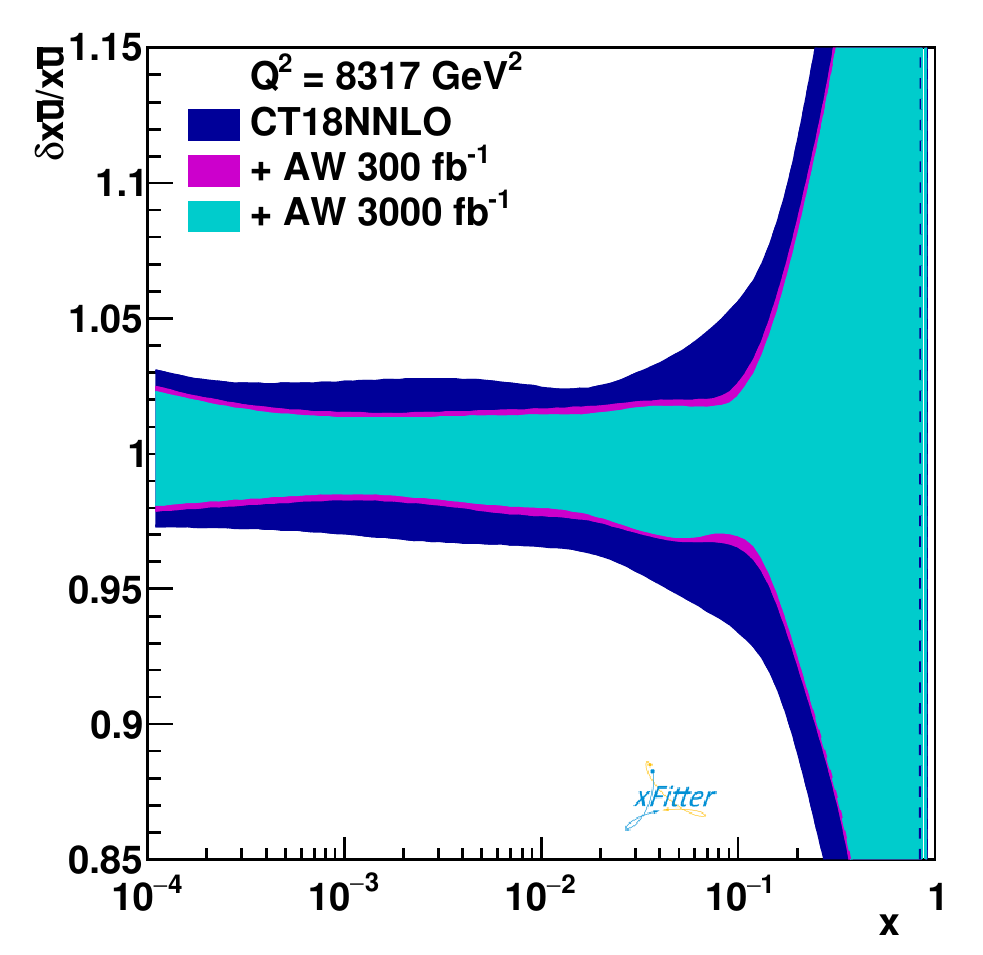}
\includegraphics[width=0.3\textwidth]{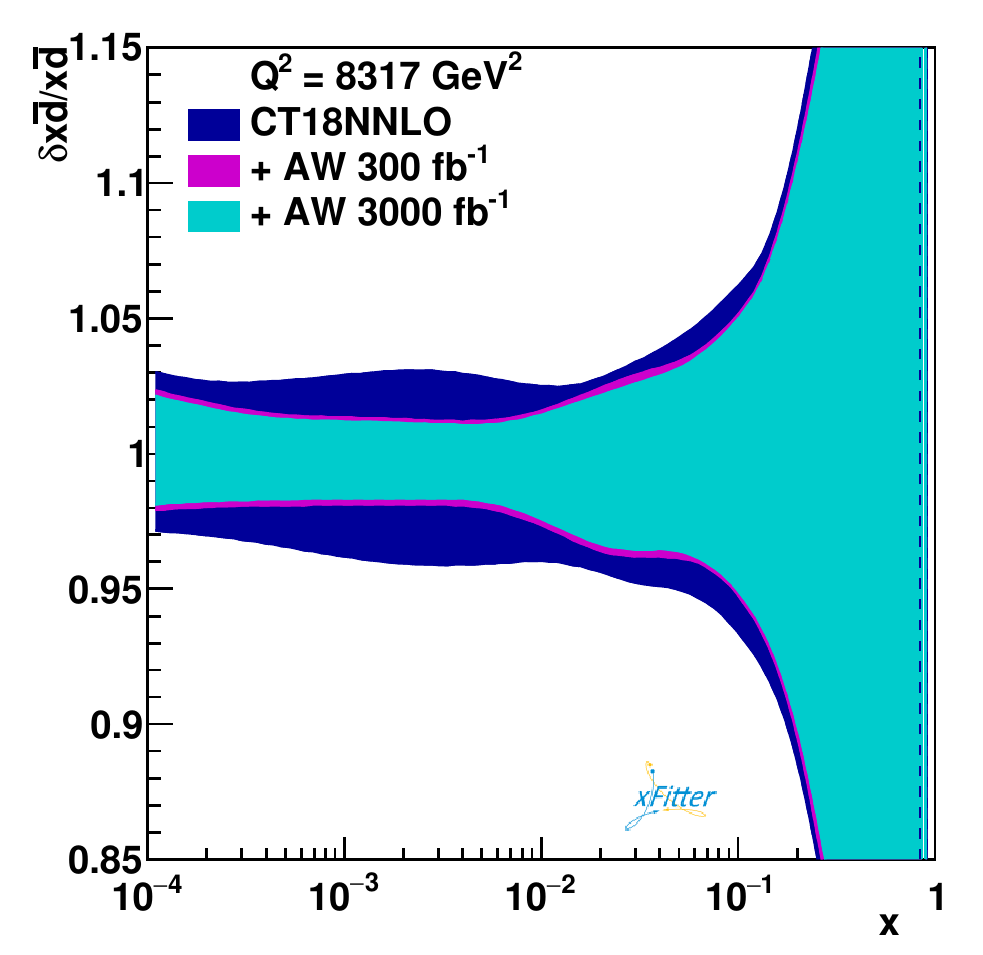}
\end{center}
\caption{Original CT18NNLO~\cite{Hou:2019efy} (blue) and profiled distributions using $A_{FB}$ (top) and $A_W$ (bottom) pseudodata corresponding to integrated luminosities of 300 fb$^{-1}$ (pink) and 3000 fb$^{-1}$ (cyan). Results are shown for anti-quark distributions.}
\label{fig:AFB_AW_antiquark}
\end{figure} 
 
Fig.~\ref{fig:AFB_AW_antiquark} shows the constraints on the anti-quark distributions and we observe that here the constraints by $A_{W}$ are slightly stronger than the ones from $A_{FB}$, particularly for the $\bar{u}$ PDF in the low $x$ region and for the $\bar{d}$ PDF in the low and intermediate $x$ range. The improvement in this case is however more moderate. 

We note that in the above results the reduction in PDF uncertainties appears to saturate with increasing luminosity, with the profiled error bands obtained for 3000 fb$^{-1}$ being close to the ones obtained for 300 fb$^{-1}$.

\subsection{Profiling with the combination of ${A_{FB}}$ and ${A_W}$}

\begin{figure}
\begin{center}
\includegraphics[width=0.3\textwidth]{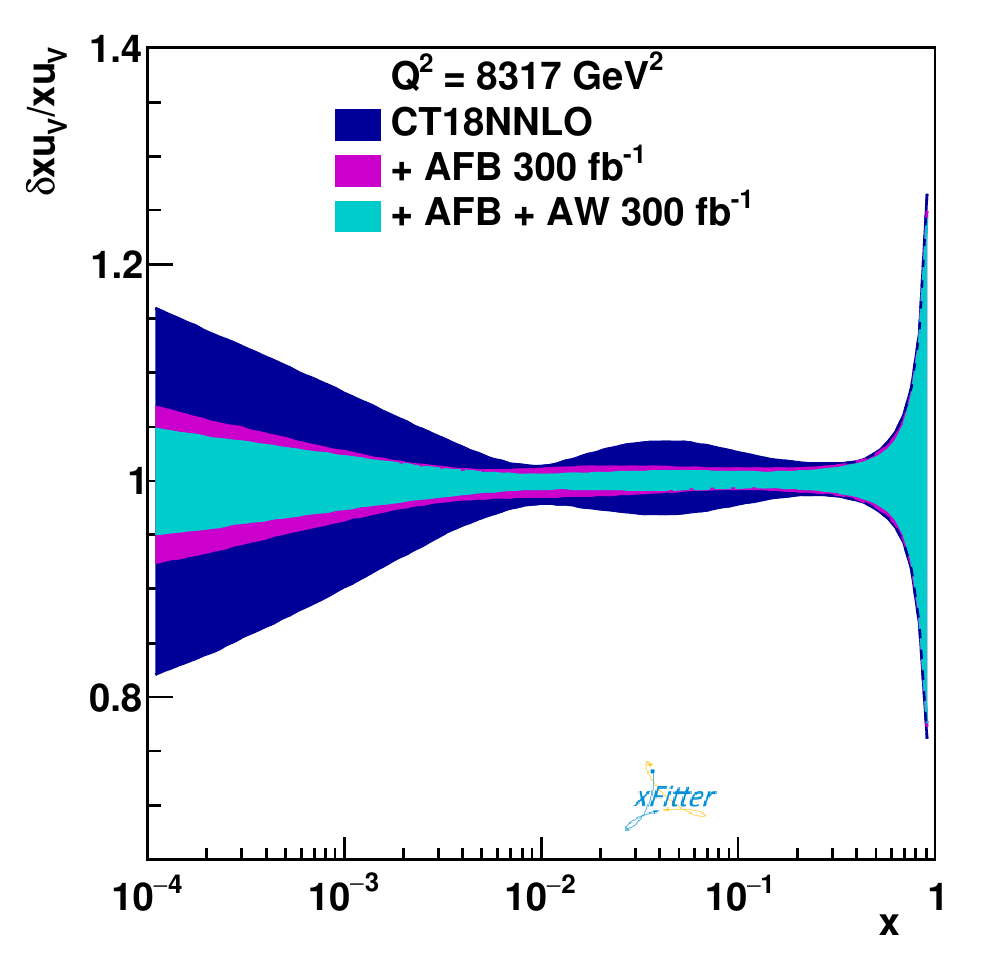}
\includegraphics[width=0.3\textwidth]{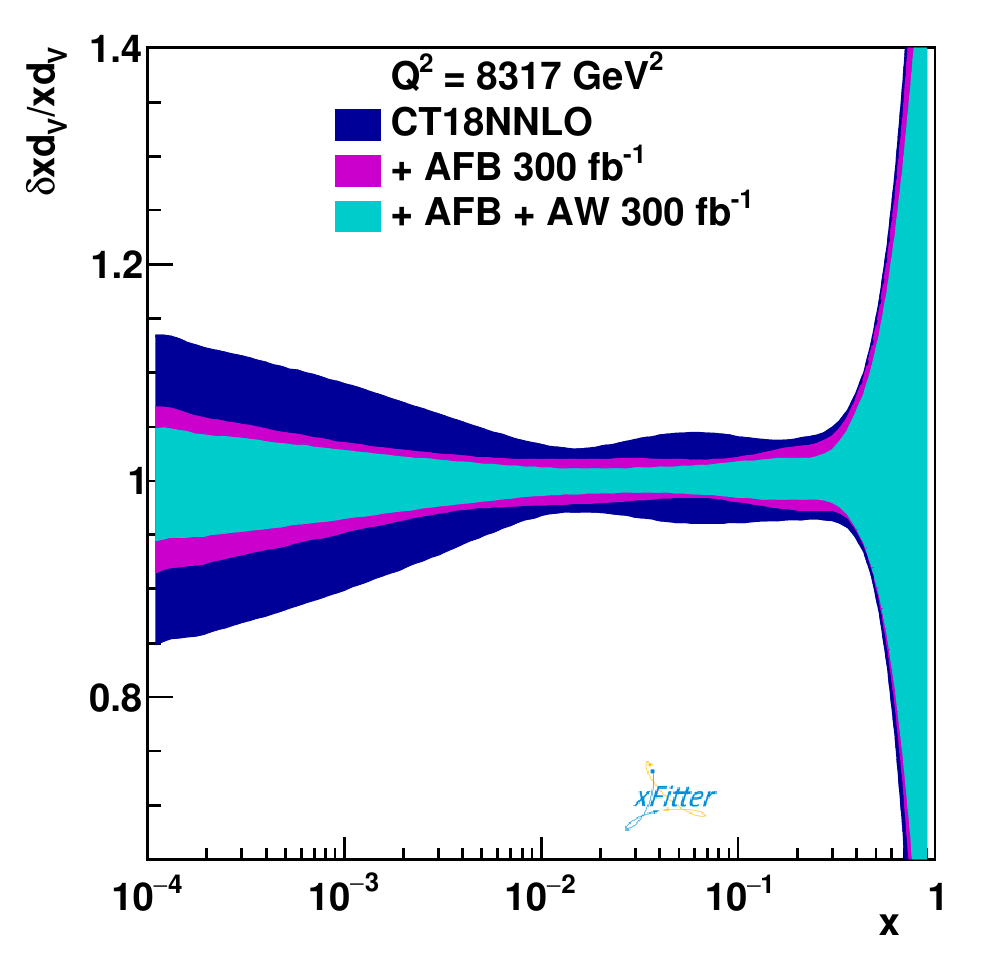}
\includegraphics[width=0.3\textwidth]{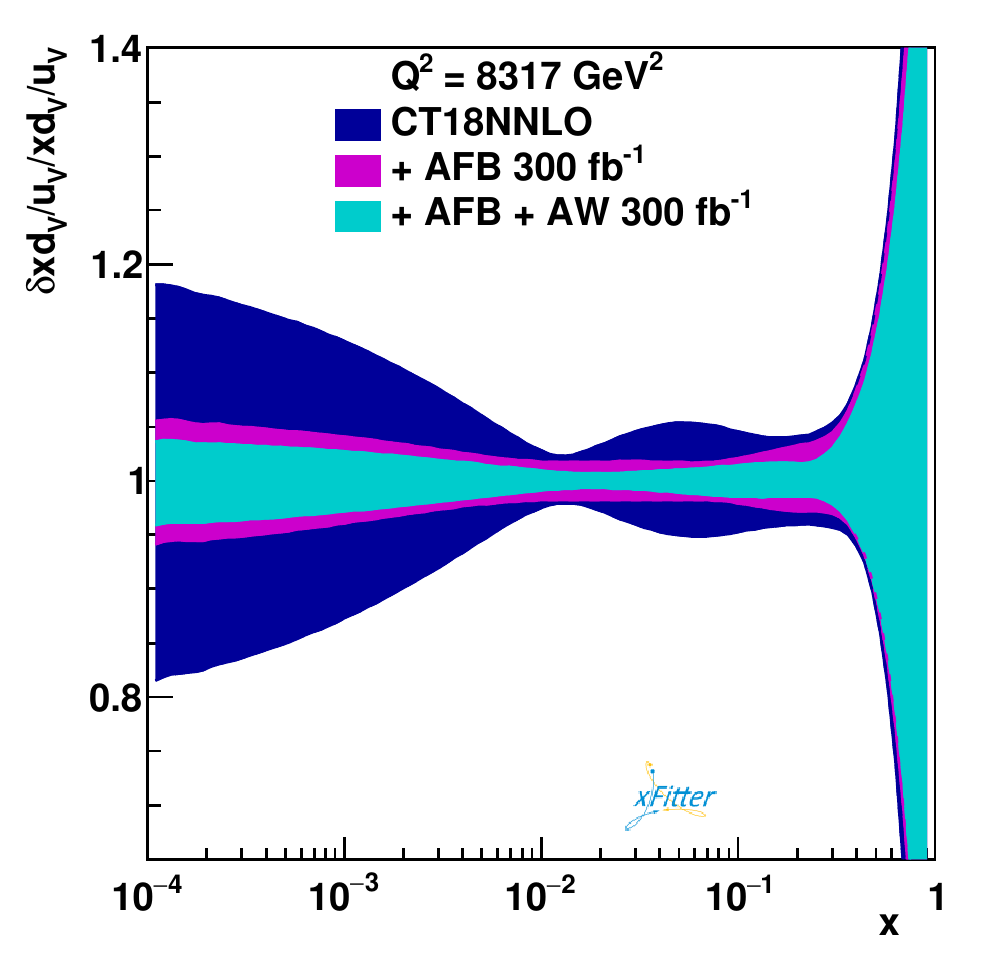}\\
\includegraphics[width=0.3\textwidth]{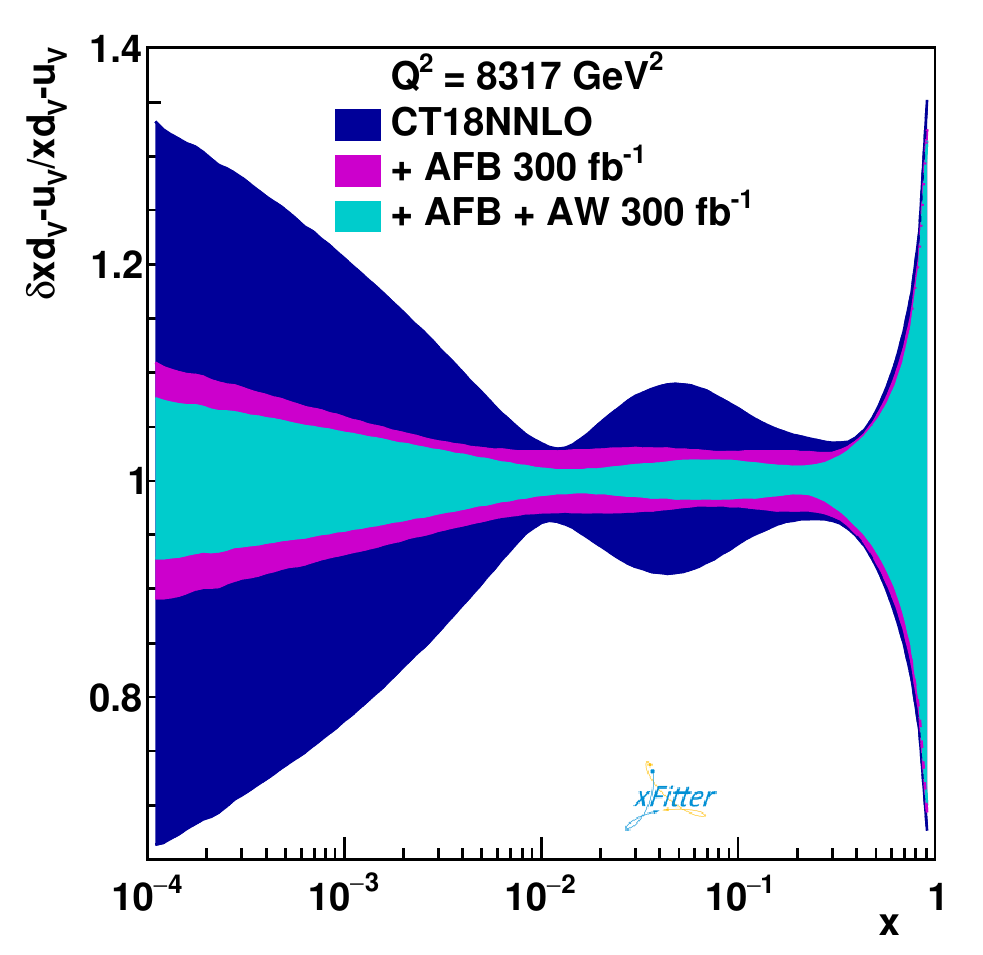}
\includegraphics[width=0.3\textwidth]{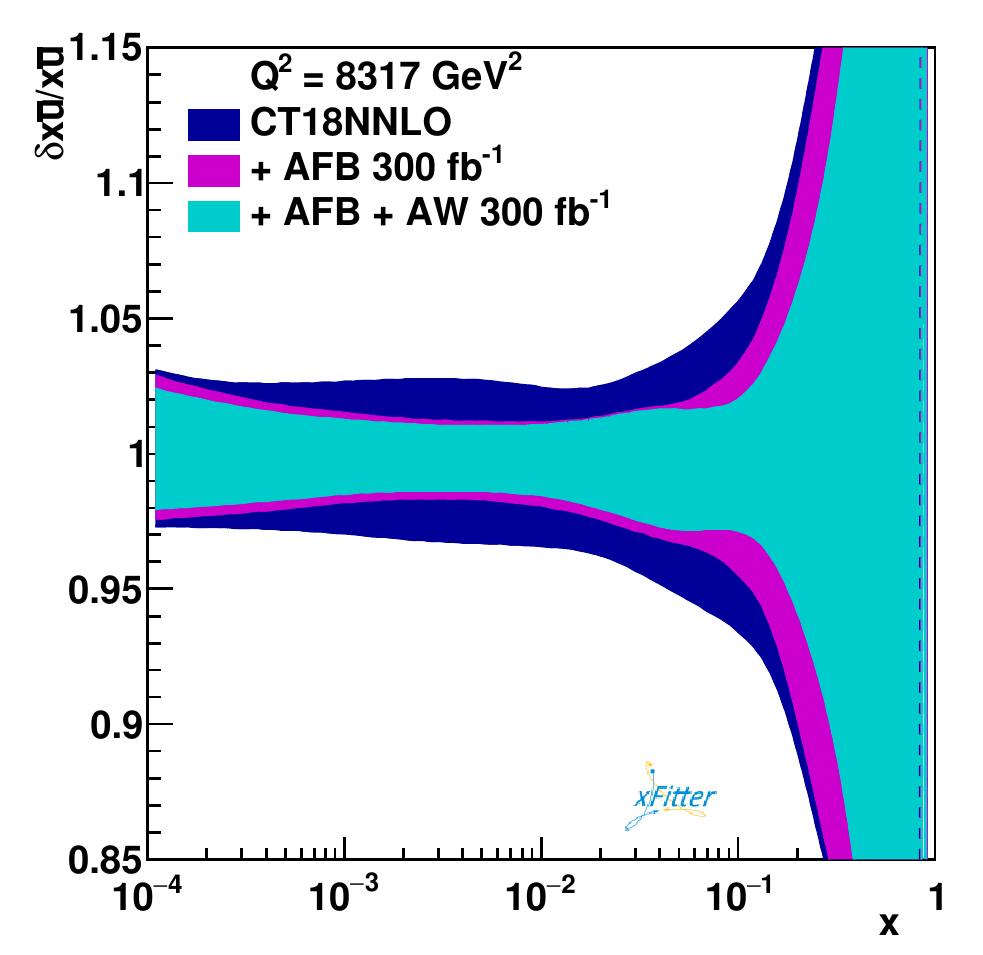}
\includegraphics[width=0.3\textwidth]{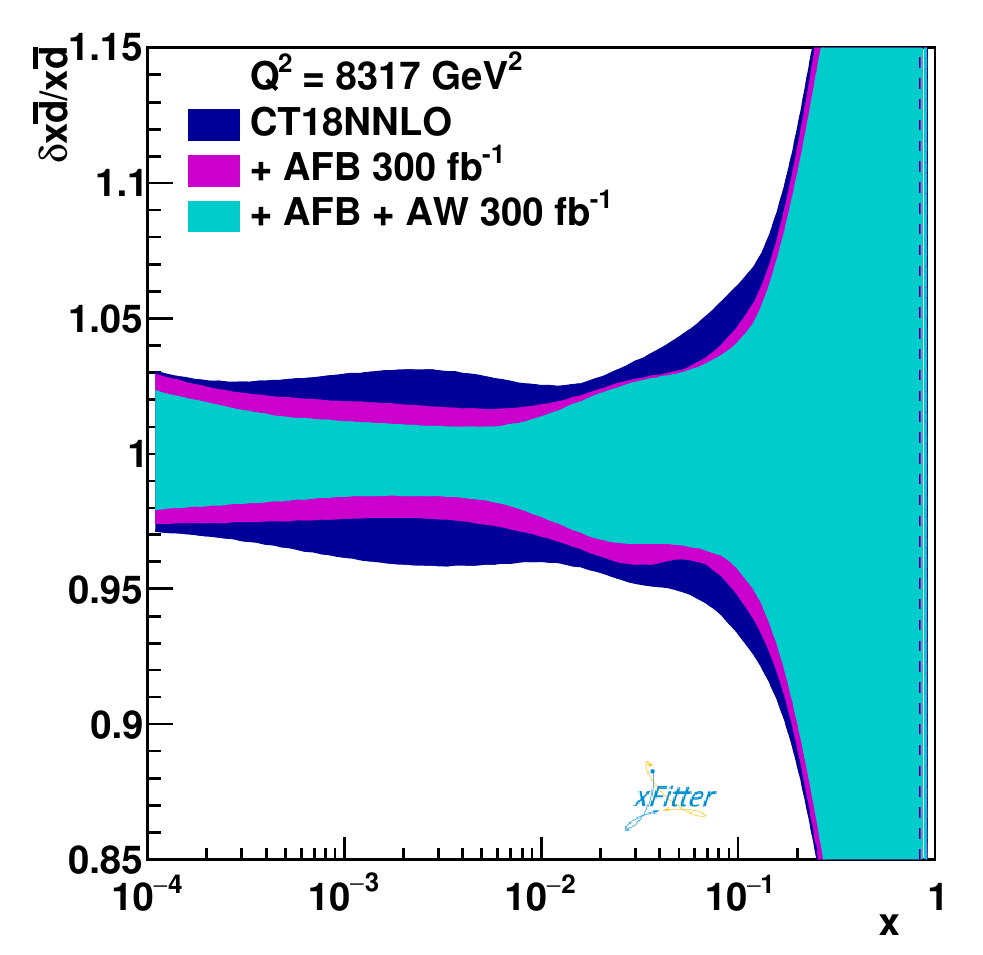}
\end{center}
\caption{Original CT18NNLO~\cite{Hou:2019efy} (blue) and profiled distributions using either $A_{FB}$ (pink) or both $A_{FB}$ and $A_W$ (cyan) pseudodata corresponding to integrated luminosity of 300 fb$^{-1}$. Results are shown for valence-quark and sea-quark distributions at $Q^2 = M_Z^2 =$ 8317 GeV$^2$.}
\label{fig:AW_AFB_comb_300}
\end{figure}

We next present the results of the profiling when superimposing constraints from $A_{FB}$ and $A_W$ pseudodata.
Fig.~\ref{fig:AW_AFB_comb_300} shows the profiled PDF set uncertainty bands using $A_{FB}$ pseudodata, and the corresponding bands upon inclusion of constraints from $A_W$ pseudodata, in the scenario of 300 fb$^{-1}$ integrated luminosity.
This illustrates how the combination of $A_{FB}$ and $A_W$ further improves PDF behaviour in terms of error reduction.
For instance, in the $d_V - u_V$ PDF combination at $x = 10^{-4}$, a 20\% reduction of uncertainty from $A_{FB}$ pseudodata is further improved by an extra 2\% by the inclusion of the $A_W$ observable.

\begin{figure}
\begin{center}
\includegraphics[width=0.3\textwidth]{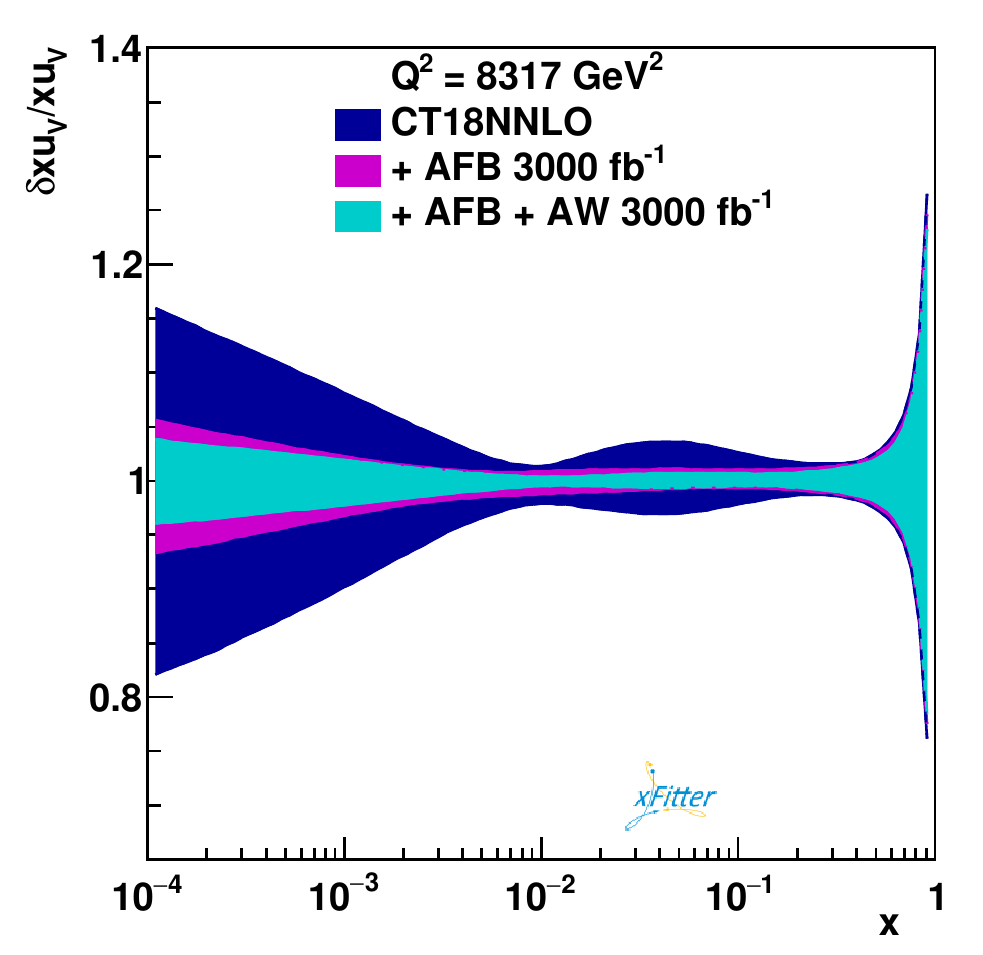}
\includegraphics[width=0.3\textwidth]{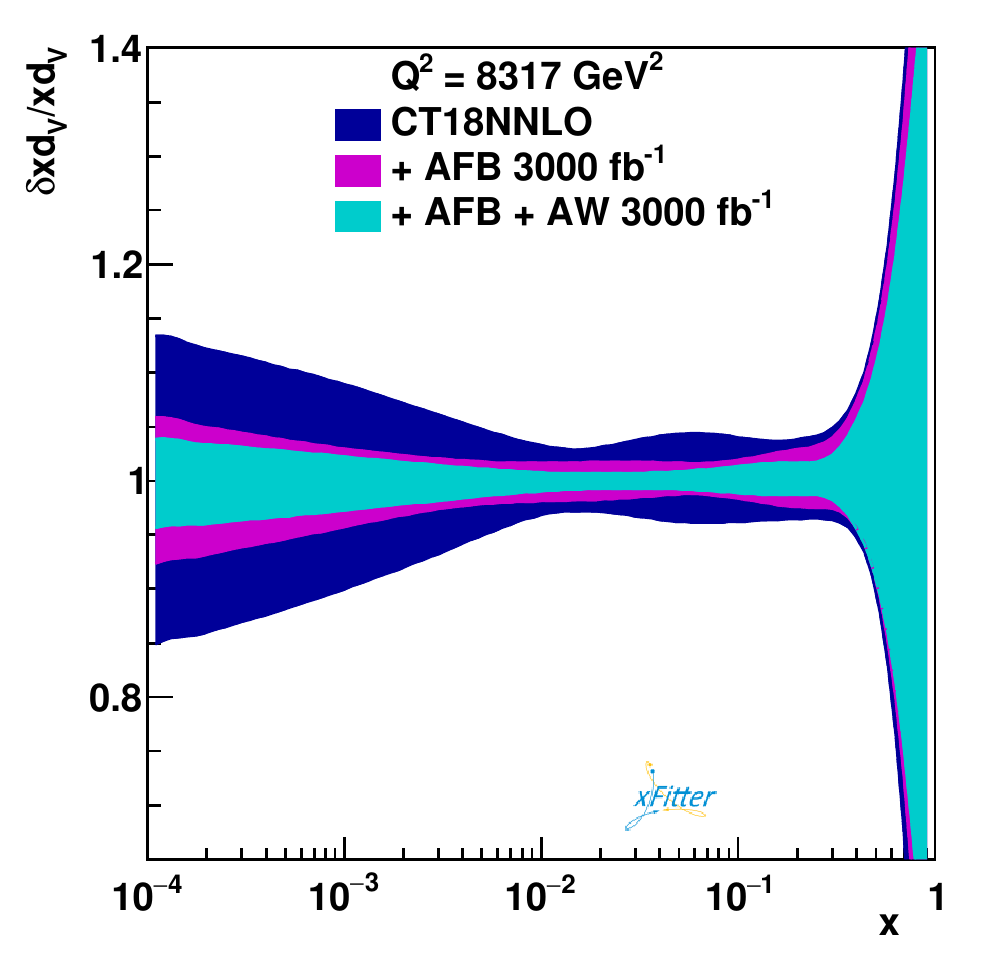}
\includegraphics[width=0.3\textwidth]{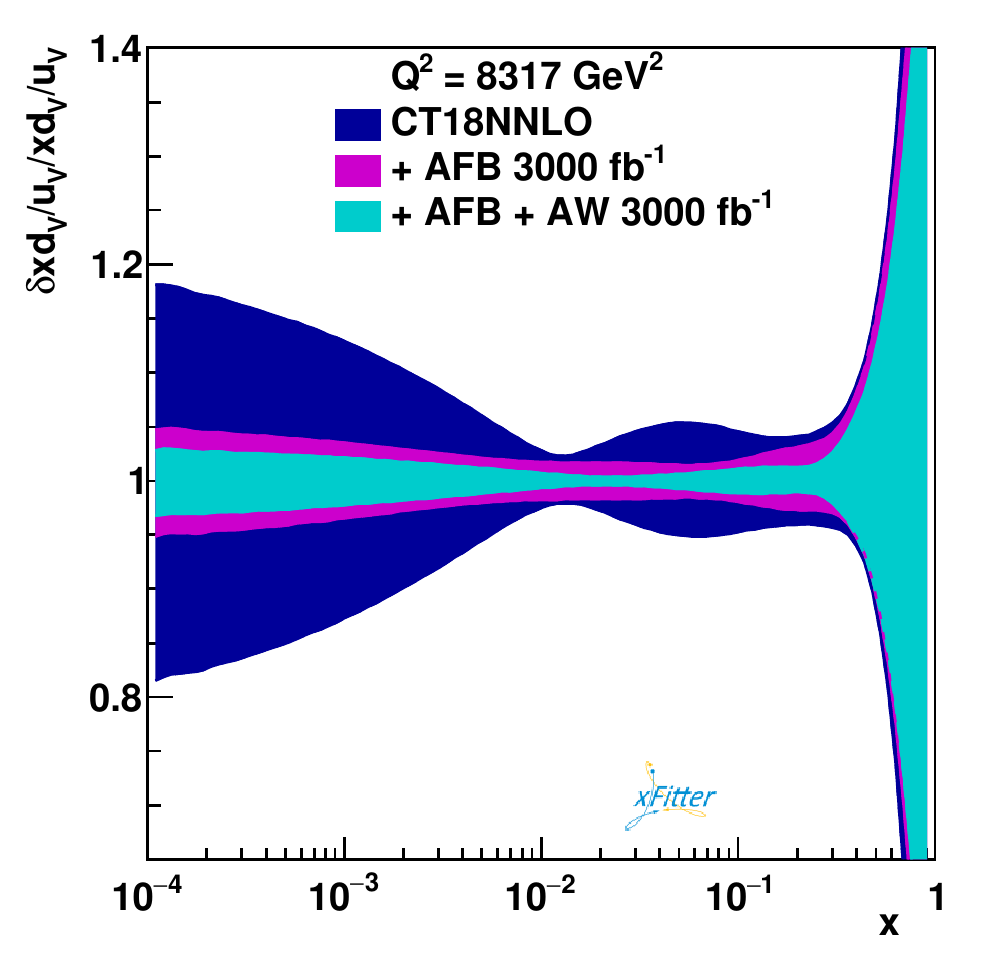}\\
\includegraphics[width=0.3\textwidth]{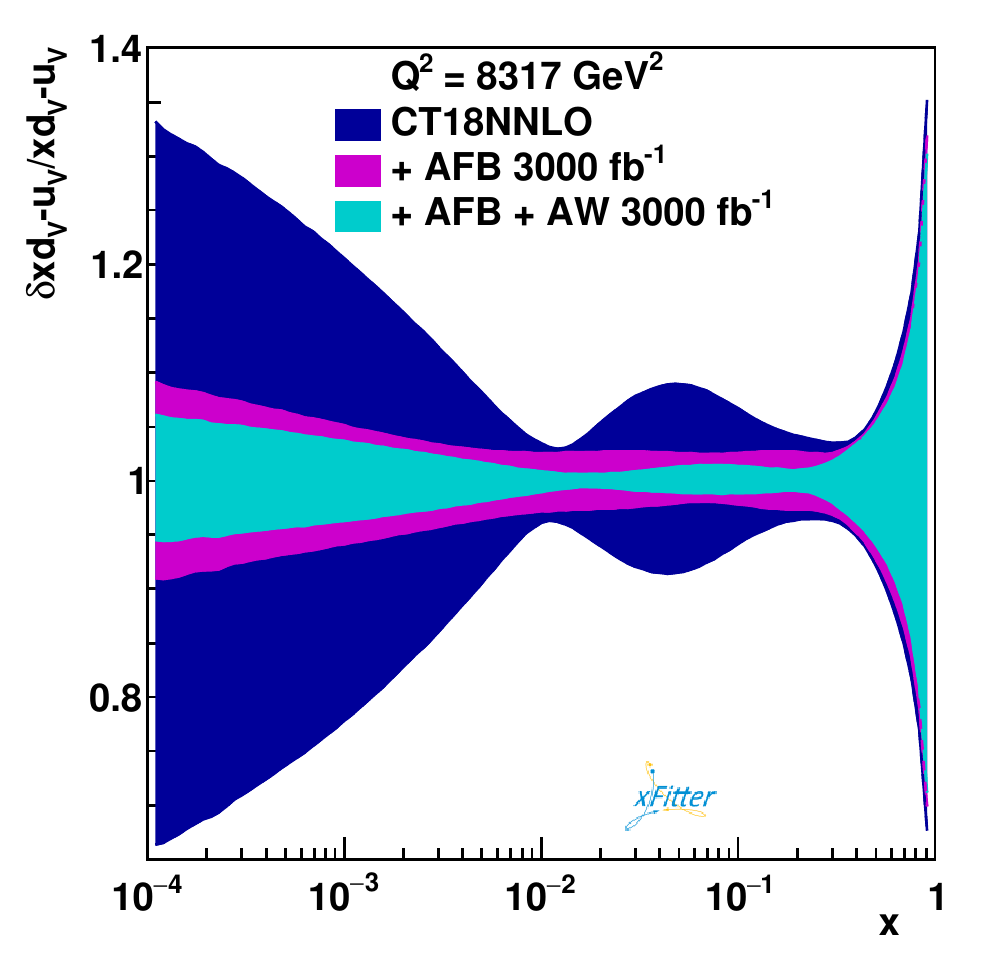}
\includegraphics[width=0.3\textwidth]{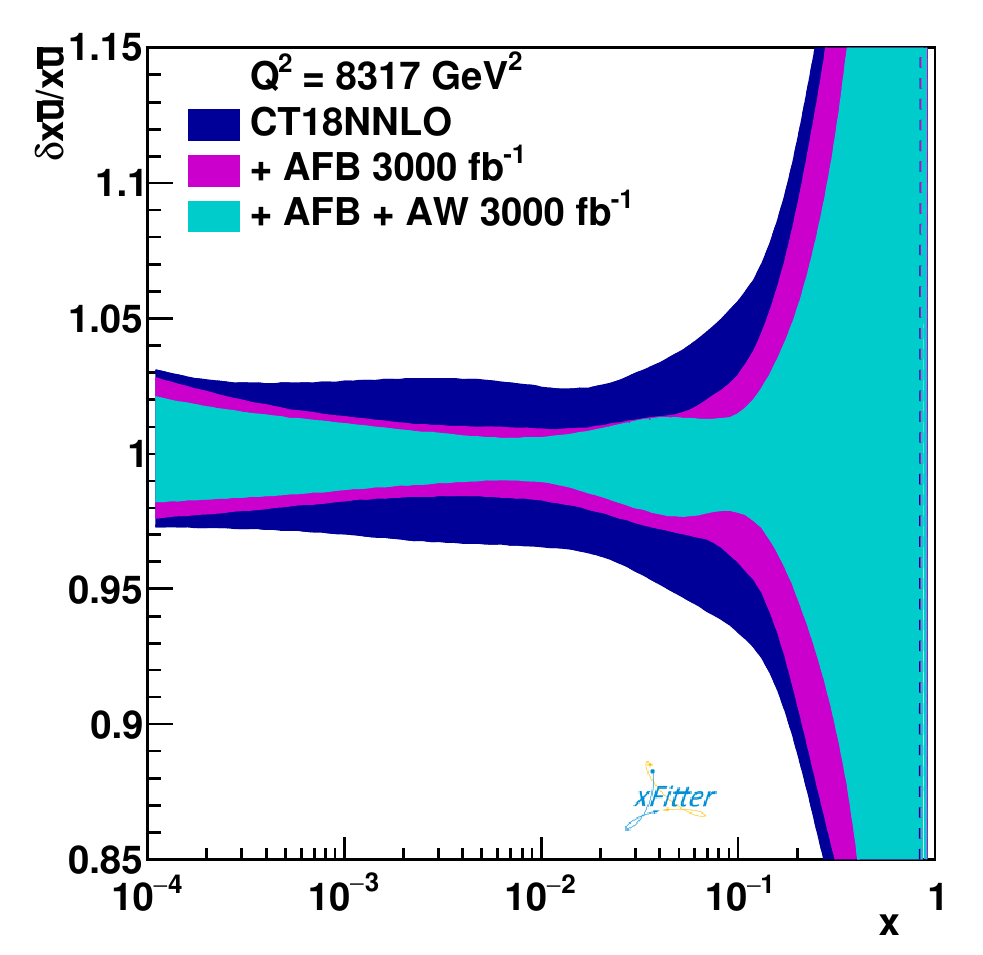}
\includegraphics[width=0.3\textwidth]{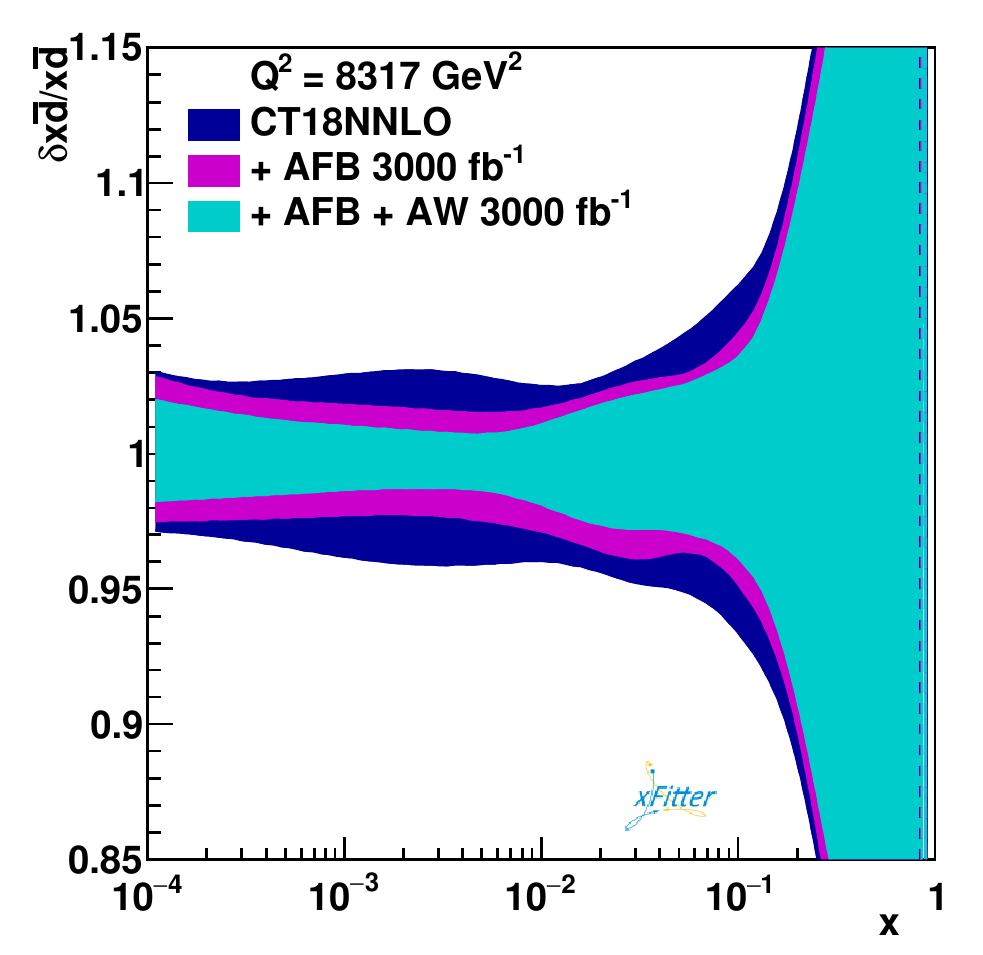}
\end{center}
\caption{Same as in Fig.~\ref{fig:AW_AFB_comb_300} but for integrated luminosity of 3000 fb$^{-1}$.}
\label{fig:AW_AFB_comb_3000}
\end{figure}

Fig.~\ref{fig:AW_AFB_comb_3000} shows the analogous results in the scenario of 3000 fb$^{-1}$ integrated luminosity.
The saturation in the reduction of uncertainties already observed in the previous subsection is also visible here.
Considering the $d_V - u_V$ PDF combination at $x = 10^{-4}$, we observe a further reduction of about 1\% with respect to the case with pseudodata corresponding to 300 fb$^{-1}$ luminosity, when superimposing $A_{FB}$ and $A_W$ constrains.

\subsection{Antimatter asymmetry in the proton}

\begin{figure}
\begin{center}
\includegraphics[width=0.3\textwidth]{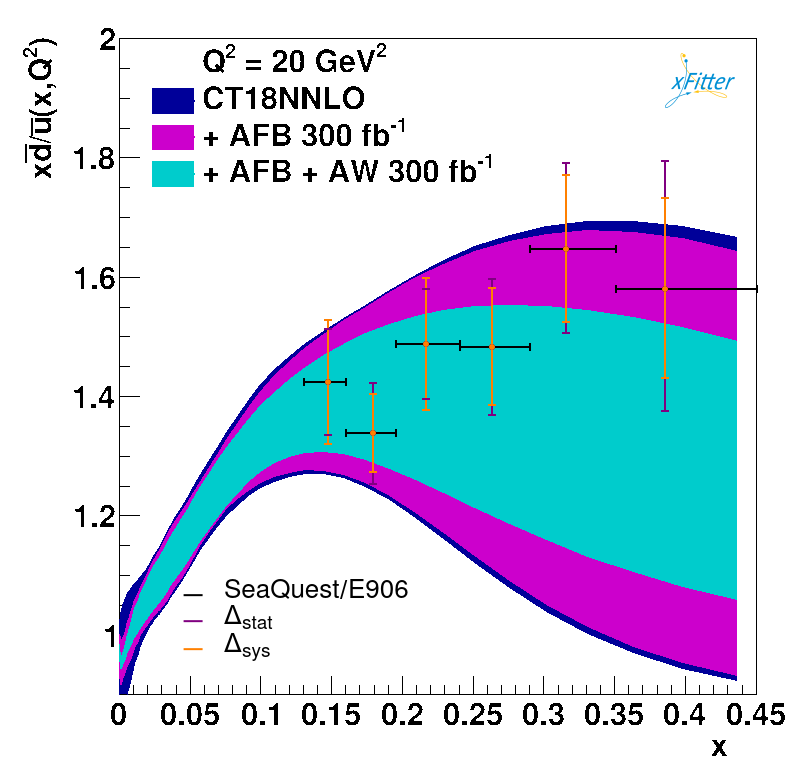}
\includegraphics[width=0.3\textwidth]{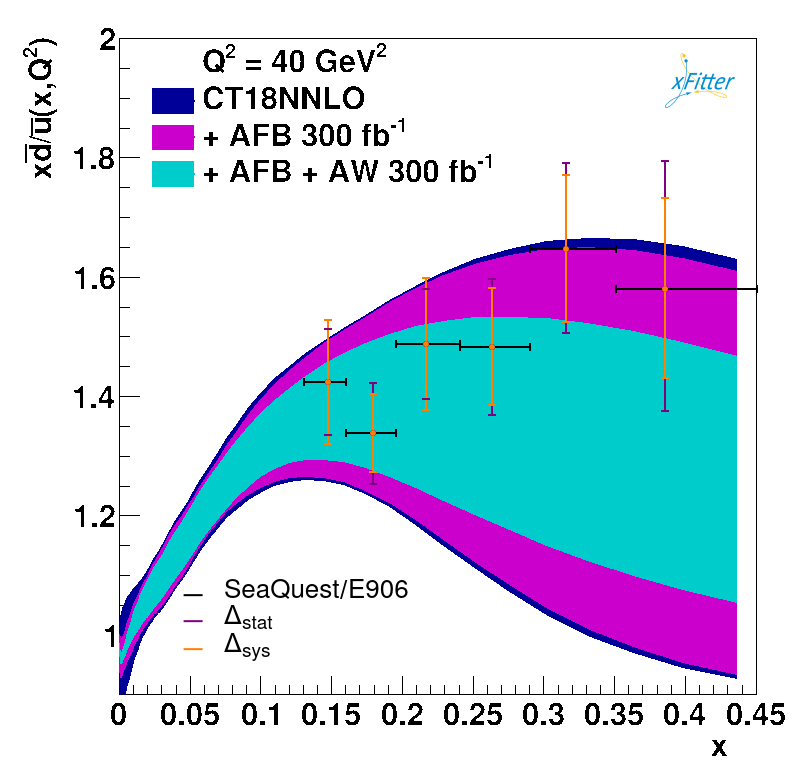}\\
\includegraphics[width=0.3\textwidth]{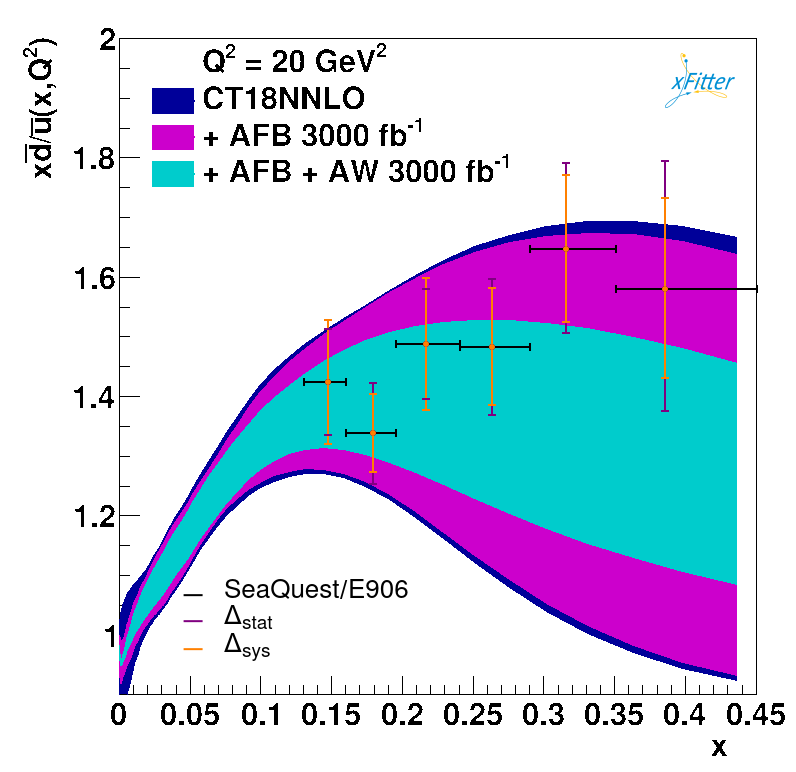}
\includegraphics[width=0.3\textwidth]{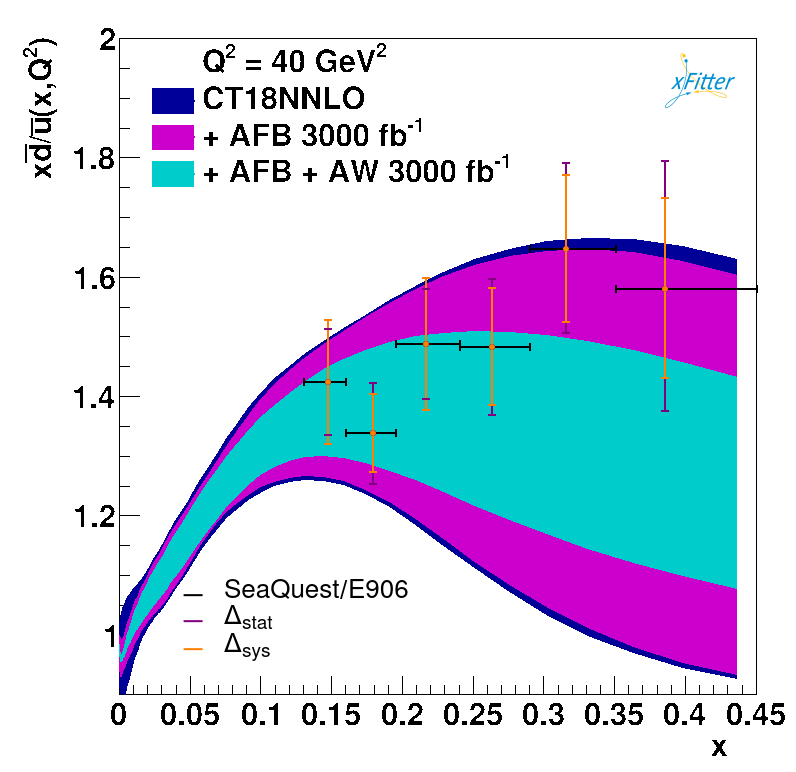}
\end{center}
\caption{Profiled error bands with $A_{FB}$ and $A_W$ asymmetries for the $\bar{d}/\bar{u}$ ratio of PDFs corresponding to 300 (top) and 3000 (bottom) fb$^{-1}$ at the energy scales $Q^2$ = 20 GeV$^2$ (left) and $Q^2$ = 40 GeV$^2$ (right). The SeaQuest/E906~\cite{Dove:2021ejl} results are superimposed for comparison.}
\label{fig:SeaQuest}
\end{figure}

In Fig.~\ref{fig:SeaQuest} we compare the improved PDF uncertainties, after profiling based on the combination of $A_{FB}$ and $A_W$ asymmetries, with the 
recent SeaQuest/E906 results~\cite{Dove:2021ejl}, wherein it was shown that the ratio of the PDFs of $\bar d$ and $\bar u$ states as a function of $x$ is notably different from both the QCD expectation of it being nearly 1~\cite{Ross:1978xk} and the predictions of several proposed mechanisms (e.g., Pauli blocking, statistical models, chiral solitons and 
meson-baryon dynamics) that had been disfavoured by similar previous results from NuSea/E866~\cite{Towell:2001nh}. The reduction in the PDF uncertainties in presence of $A_{FB}$ and (especially) $A_W$ constraints is significant, of up to a factor 2 in the very high $x$ region, to the extent that data over the latter are no longer within the PDF errors, no matter the actual values of $Q^2$ and luminosity considered. 
We remark that however in order to produce these results we have employed pseudodata projecting future LHC run statistics. The inclusion of future real data would likely modify the central values of the PDF distributions as well.

\section{DY production at the EW mass scale and at TeV masses}

In this section we present examples illustrating the implications of the analysis in Sec.~\ref{sec:compl} for DY observables both in the mass region near the SM vector-boson masses and in the multi-TeV mass region relevant for BSM searches. 

\subsection{PDF uncertainties in dilepton observables}
We here examine theoretical uncertainties on dilepton observables due to the original PDF sets and the profiled sets obtained after imposing constraints from $A_{FB}$ and $A_W$ asymmetry measurements both with 300 fb$^{-1}$ and 3000 fb$^{-1}$ of luminosity.

While the best $Z$-boson mass determinations come from $e^+ e^-$ experimental data, the LHC provides competitive determinations of $W$-boson mass~\cite{Aaboud:2017svj,ATLAS:2018qzr}. Motivated by this, we start by considering the transverse mass spectrum and lepton transverse momentum spectrum of the charged DY channel.

\begin{figure}
\begin{center}
\includegraphics[width=0.34\textwidth]{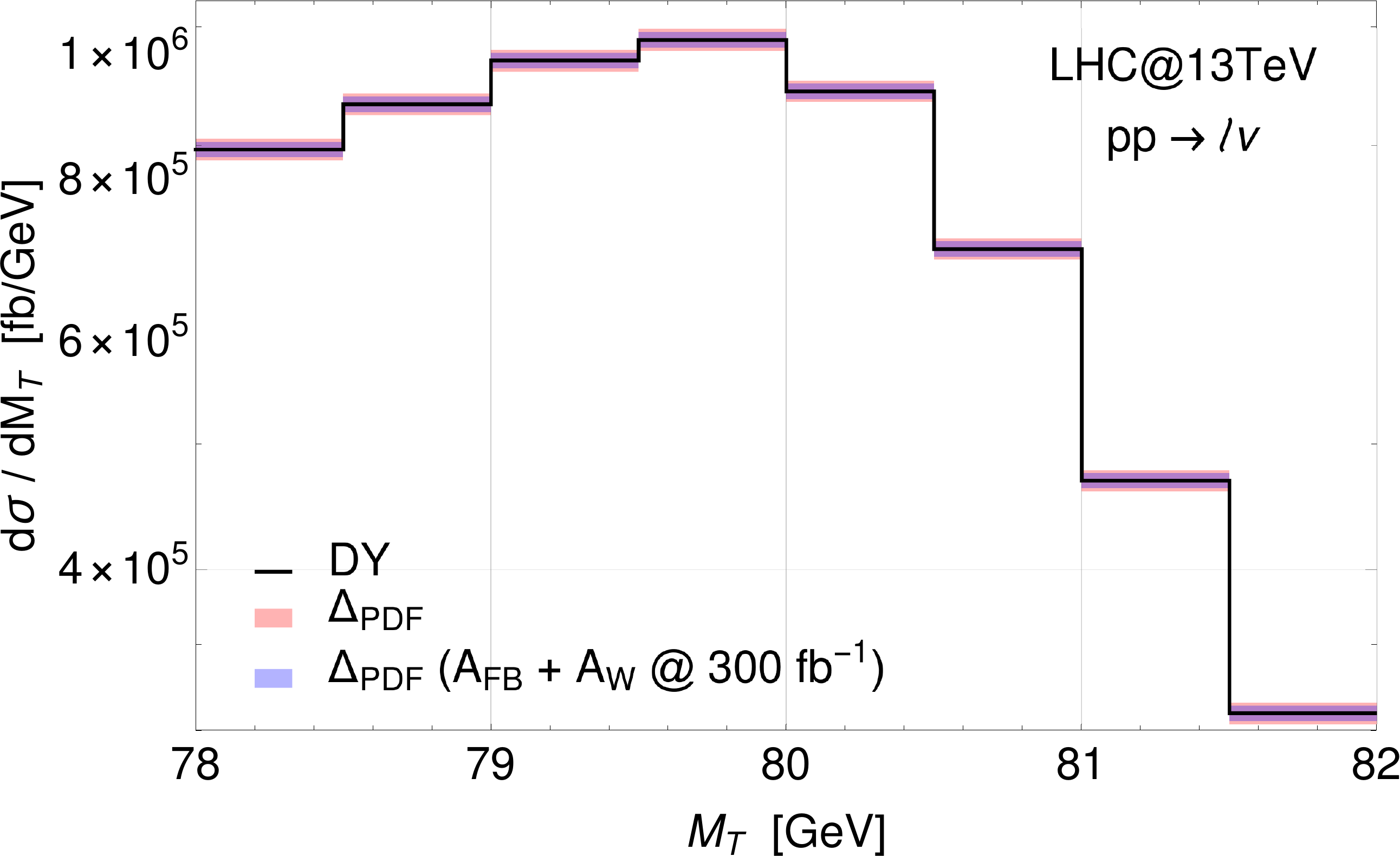}
\includegraphics[width=0.5\textwidth]{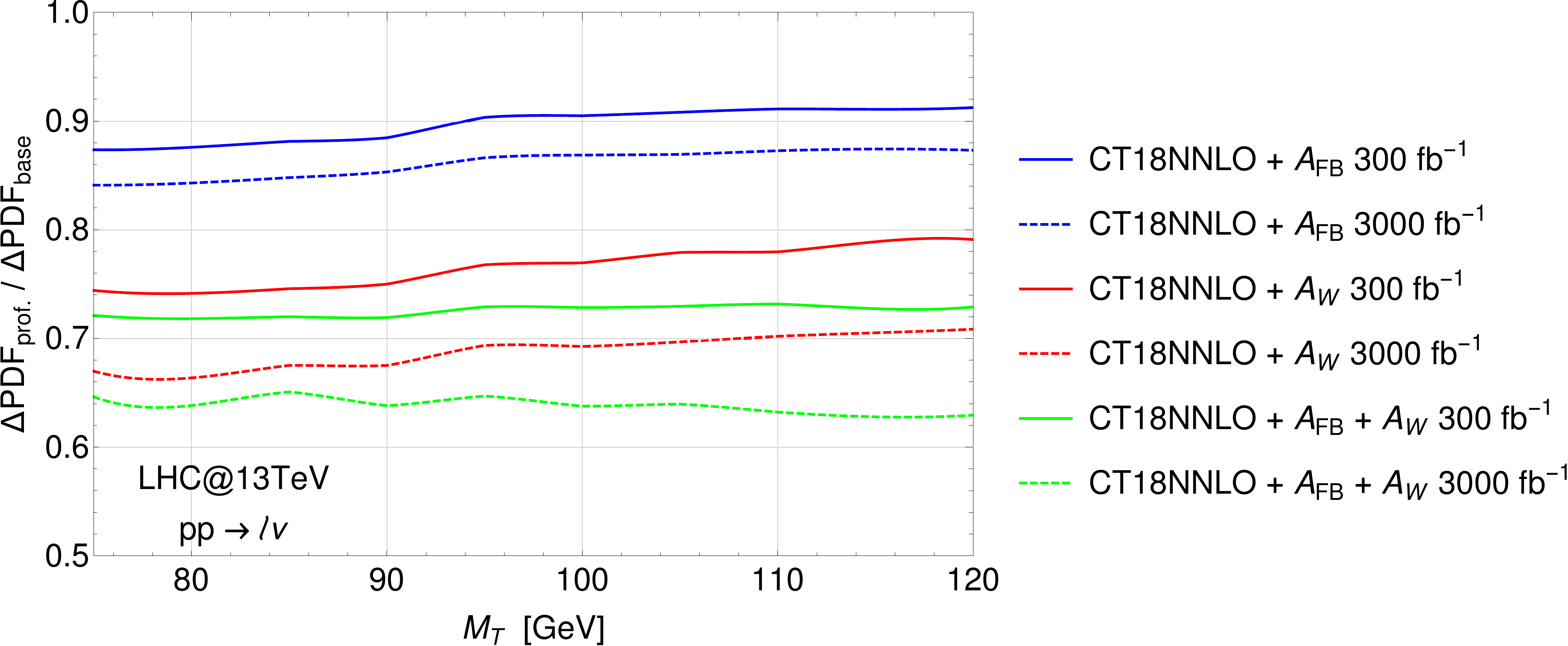}
\end{center}
\caption{(left) Charged-current DY transverse mass distribution with PDF error; (right) relative improvement of PDF error on the transverse mass spectrum due to profiling based on $A_{FB}$, $A_W$ and their combination.}
\label{fig:MT_PDF_error}
\end{figure}

Fig.~\ref{fig:MT_PDF_error} on the left shows the charged-current DY transverse mass spectrum with the relative PDF uncertainty bands obtained by the baseline CT18NNLO (rescaled at 68\% CL) and by the profiled PDF set using the combination of $A_{FB}$ and $A_W$ measurements corresponding to 300 fb$^{-1}$ of integrated luminosity, while on the right the ratio is shown for the profiled PDF error over the original one.
The original PDF uncertainty in this region varies between 1.7\% and 1.9\%. The inclusion of $A_{FB}$ constraints at 300 (3000) fb$^{-1}$ reduces the uncertainty by about 12\% (16\%) while the inclusion of $A_{W}$ constraints at 300 (3000) fb$^{-1}$ reduces the uncertainty by about 26\% (43\%). By combining the two sets of constraints 
at 300 (3000) fb$^{-1}$ the original PDF error bands are reduced by about 28\% (46\%).

\begin{figure}
\begin{center}
\includegraphics[width=0.34\textwidth]{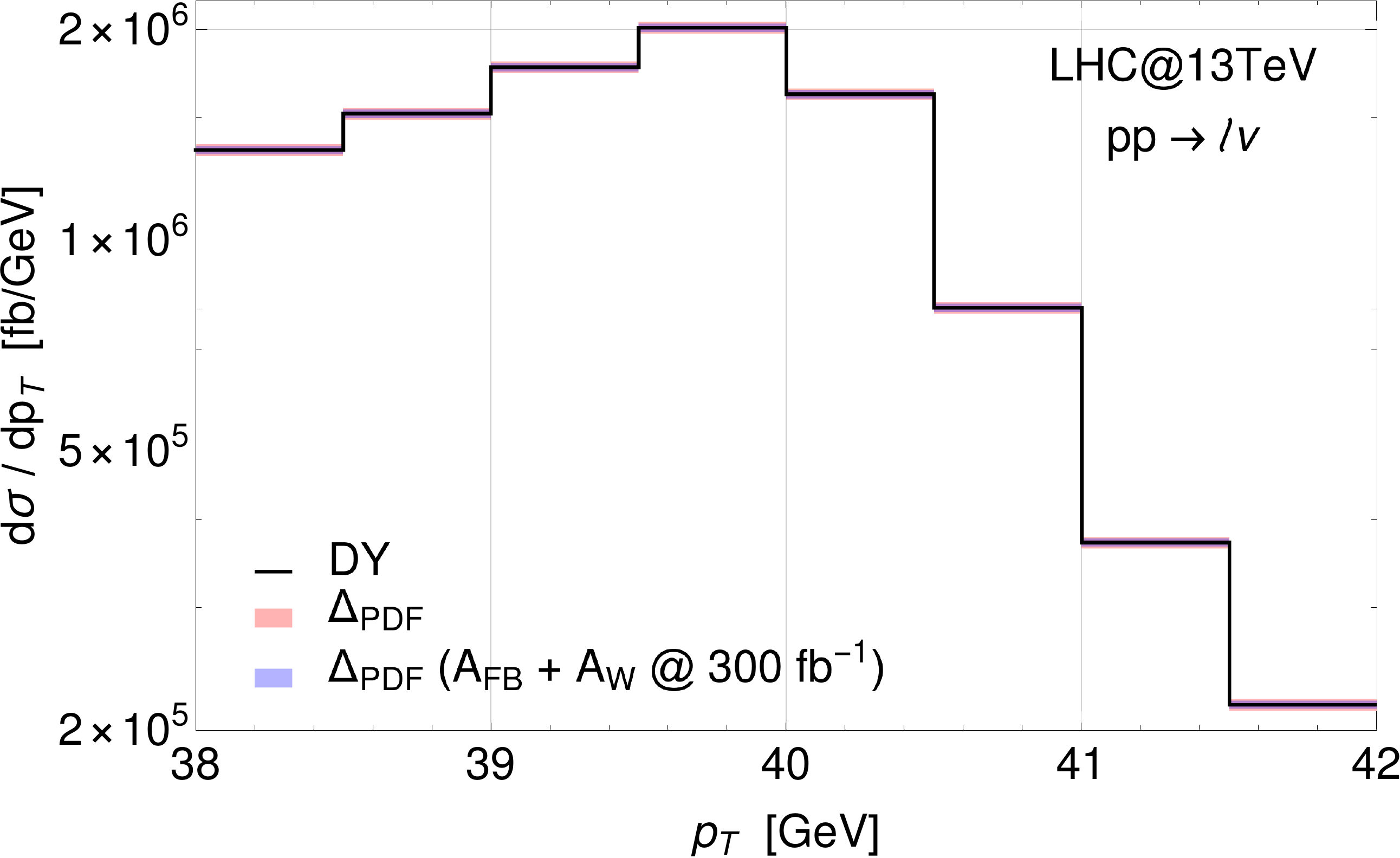}
\includegraphics[width=0.5\textwidth]{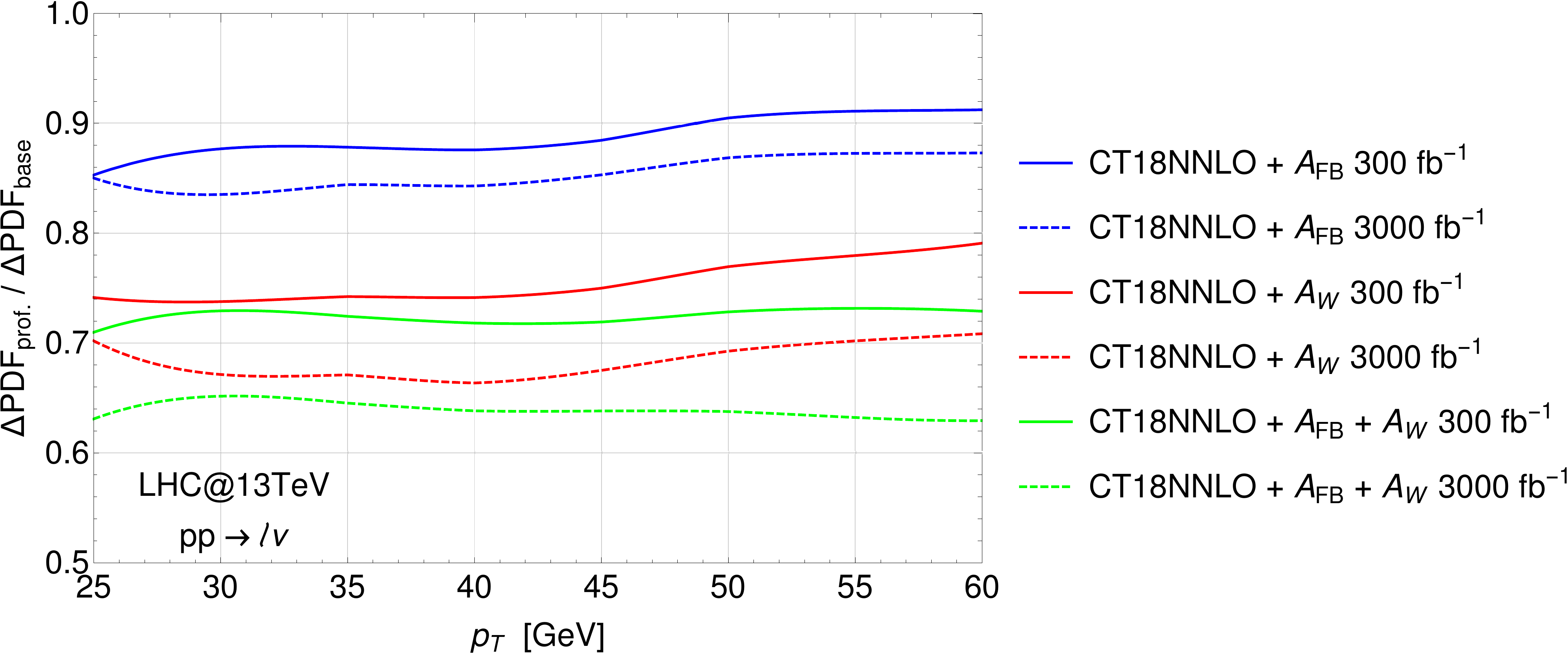}
\end{center}
\caption{(left) Lepton transverse momentum distribution with PDF error; (right) relative improvement of PDF error on the lepton transverse momentum spectrum 
due to profiling based on $A_{FB}$, $A_W$ and their combination.}
\label{fig:pT_PDF_error}
\end{figure}

Fig.~\ref{fig:pT_PDF_error} shows analogous results for the lepton transverse momentum spectrum. Again the original CT18NNLO (rescaled at 68\% CL) PDF uncertainty ranges between 1.7\% and 1.9\% and its reduction obtained via asymmetry profiling has a similar behaviour to that obtained for the transverse mass distribution.

\begin{figure}
\begin{center}
\includegraphics[width=0.49\textwidth]{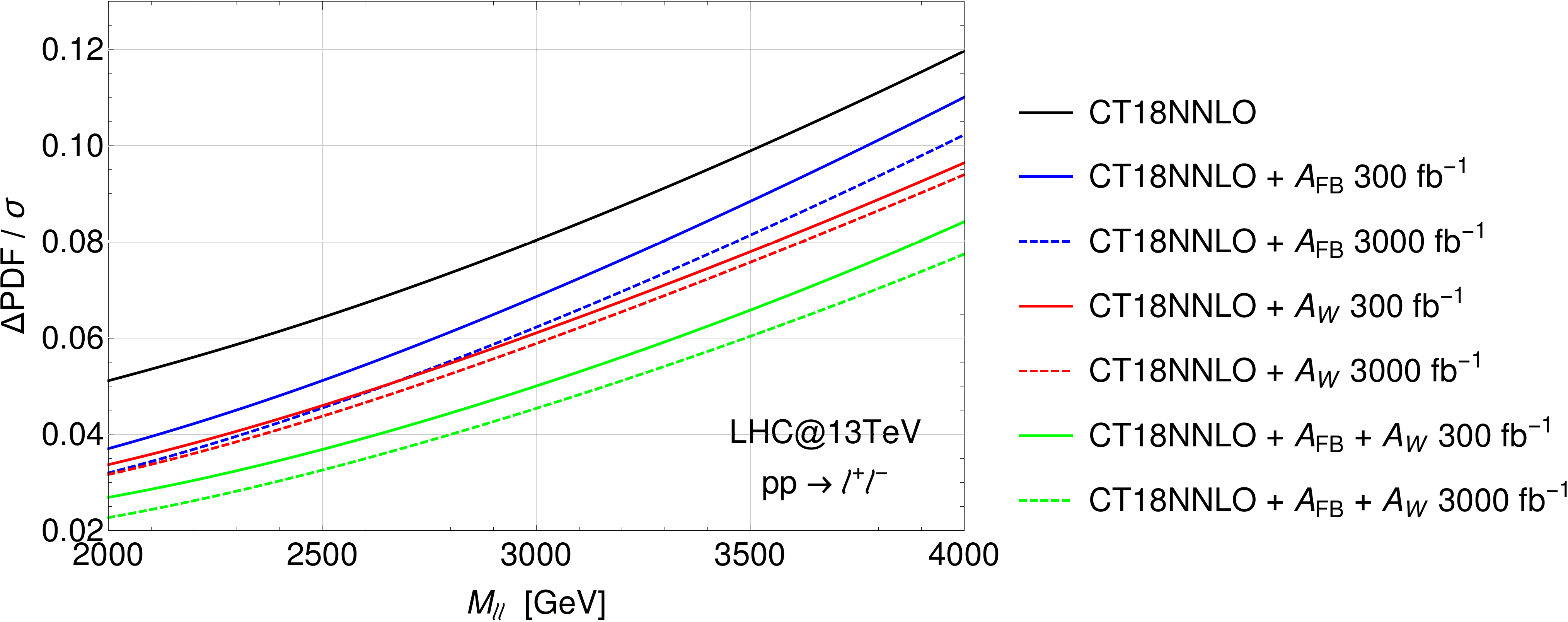}
\includegraphics[width=0.49\textwidth]{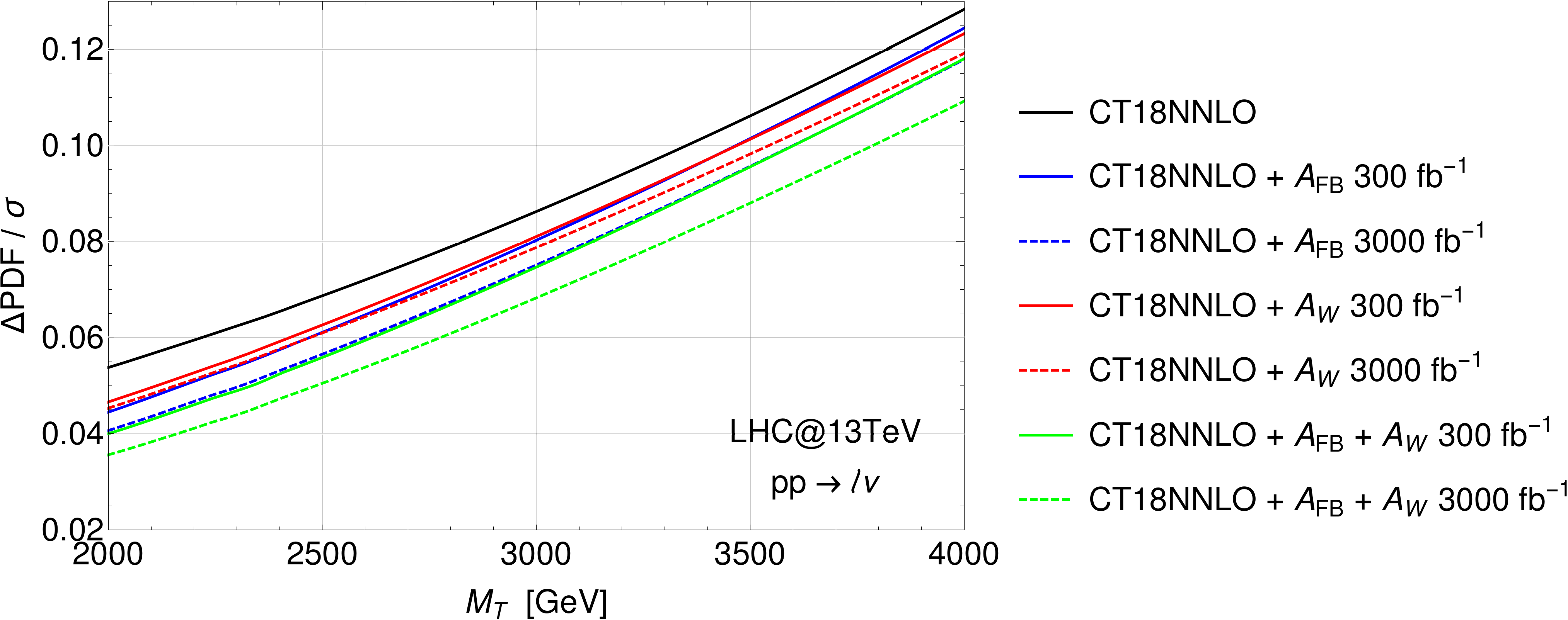}
\end{center}
\caption{Relative PDF error on the invariant mass spectrum of the neutral DY channel (left) and on the transverse mass spectrum of the charged DY channel (right).}
\label{fig:DY_PDF_error}
\end{figure}

We next consider applications of our studies to the multi-TeV region, relevant for new physics searches, e.g, for new $Z^\prime$ and $W^\prime$ 
heavy boson states.
In Fig.~\ref{fig:DY_PDF_error} on the left we can observe the relative PDF error in the dilepton invariant mass distribution.
The black curve represents the baseline CT18NNLO uncertainty while the coloured curves are the PDF errors after the profiling using the $A_{FB}$ (blue curves) and $A_W$ (red curves) measurements and their combination (green curves), corresponding to integrated luminosities of 300 fb$^{-1}$ (solid curves) and 3000 fb$^{-1}$ (dashed curves).
$A_{FB}$ reduces the PDF relative uncertainty from 5\% to 3.8\% (3.2\%) for an invariant mass of 2 TeV and from 12\% to 11\% (10.2\%) for an invariant mass of 4 TeV when using an integrated luminosity of 300 (3000) fb$^{-1}$.
$A_W$ is able to reduce the PDF relative uncertainty to 3.4\% (3.2\%) for an invariant mass of 2 TeV and to 9.6\% (9.4\%) for an invariant mass of 4 TeV when using an integrated luminosity of 300 (3000) fb$^{-1}$.
The combination of $A_{FB}$ and $A_W$ further constrains the PDF relative uncertainty to 2.7\% (2.3\%) for an invariant mass of 2 TeV and to 8.4\% (7.8\%) for an invariant mass of 4 TeV when using an integrated luminosity of 300 (3000) fb$^{-1}$.

In Fig.~\ref{fig:DY_PDF_error} on the right, the analogous analysis is presented for the charged DY channel in the transverse mass spectrum.
The black curve represents the baseline CT18NNLO uncertainty while the coloured curves are the PDF errors after the profiling using the $A_{FB}$ (blue curves) and $A_W$ (red curves) measurements and their combination (green curves), corresponding to integrated luminosities of 300 fb$^{-1}$ (solid curves) and 3000 fb$^{-1}$ (dashed curves).
$A_{FB}$ reduces the PDF relative uncertainty from 5.4\% to 4.5\% (4.1\%) for a transverse mass of 2 TeV and from 12.9\% to 12.5\% (11.8\%) for a transverse mass of 4 TeV when using an integrated luminosity of 300 (3000) fb$^{-1}$.
$A_W$ is able to reduce the PDF relative uncertainty to 4.7\% (4.6\%) for a transverse mass of 2 TeV and to 12.3\% (11.9\%) for a transverse mass of 4 TeV when using an integrated luminosity of 300 (3000) fb$^{-1}$.
The combination of $A_{FB}$ and $A_W$ data further constrains the PDF relative uncertainty to 4.0\% (3.6\%) for a transverse mass of 2 TeV and to 11.8\% (10.9\%) for a transverse mass of 4 TeV when using an integrated luminosity of 300 (3000) fb$^{-1}$.

In the following we will study the impact of the reduction of PDF uncertainties in the invariant mass and transverse mass spectra of specific BSM benchmarks featuring a neutral and charged resonance, respectively.
We introduce the enhanced Sequential SM (SSM) model as in Ref.~\cite{Accomando:2019ahs}, which follows the original SSM~\cite{Altarelli:1989ff} and features in its spectrum an extra $Z^\prime$ and a $W^\prime$ characterised by the same chiral couplings as the SM gauge bosons, but in this case the overall BSM gauge coupling is enhanced by a factor 3 with respect to the SM one.
This is a BSM scenario providing typically wide $Z^\prime$ ($\Gamma_{Z^\prime} / M_{Z^\prime} \simeq$ 27\%) and $W^\prime$ ($\Gamma_{W^\prime} / M_{W^\prime} \simeq$ 41.6\%) whose discovery will occur through (low mass) tail effects rather than on-peak effects. 
In the following analysis we fix the masses of the neutral and charged heavy gauge bosons to 7.2 TeV and 10 TeV respectively. We will show that the reduction of PDF uncertainties is relevant for an early discovery of their associated signals at the HL-LHC, i.e., for a default luminosity of 3000 fb$^{-1}$.

\subsection{Effects on $Z^\prime$ searches}

The anticipated improvement of the PDFs in the invariant mass spectrum have a substantial impact on searches for BSM neutral resonances.
Traditional searches for peaked Breit-Wigner shapes are clearly less concerned by the errors on the PDFs, however, the significance of wide resonance signals is greatly affected by PDF systematic uncertainties.
In particular, it has been shown that the presence of a heavy broad resonance can be detected through its interference with the SM background in the invariant mass region below the $Z^\prime$ mass~\cite{Accomando:2019ahs}. Thus, the improvement of the PDF uncertainties in the relevant invariant mass region would considerably enhance the significance of a signal of this kind.

\begin{figure}
\begin{center}
\includegraphics[width=0.49\textwidth]{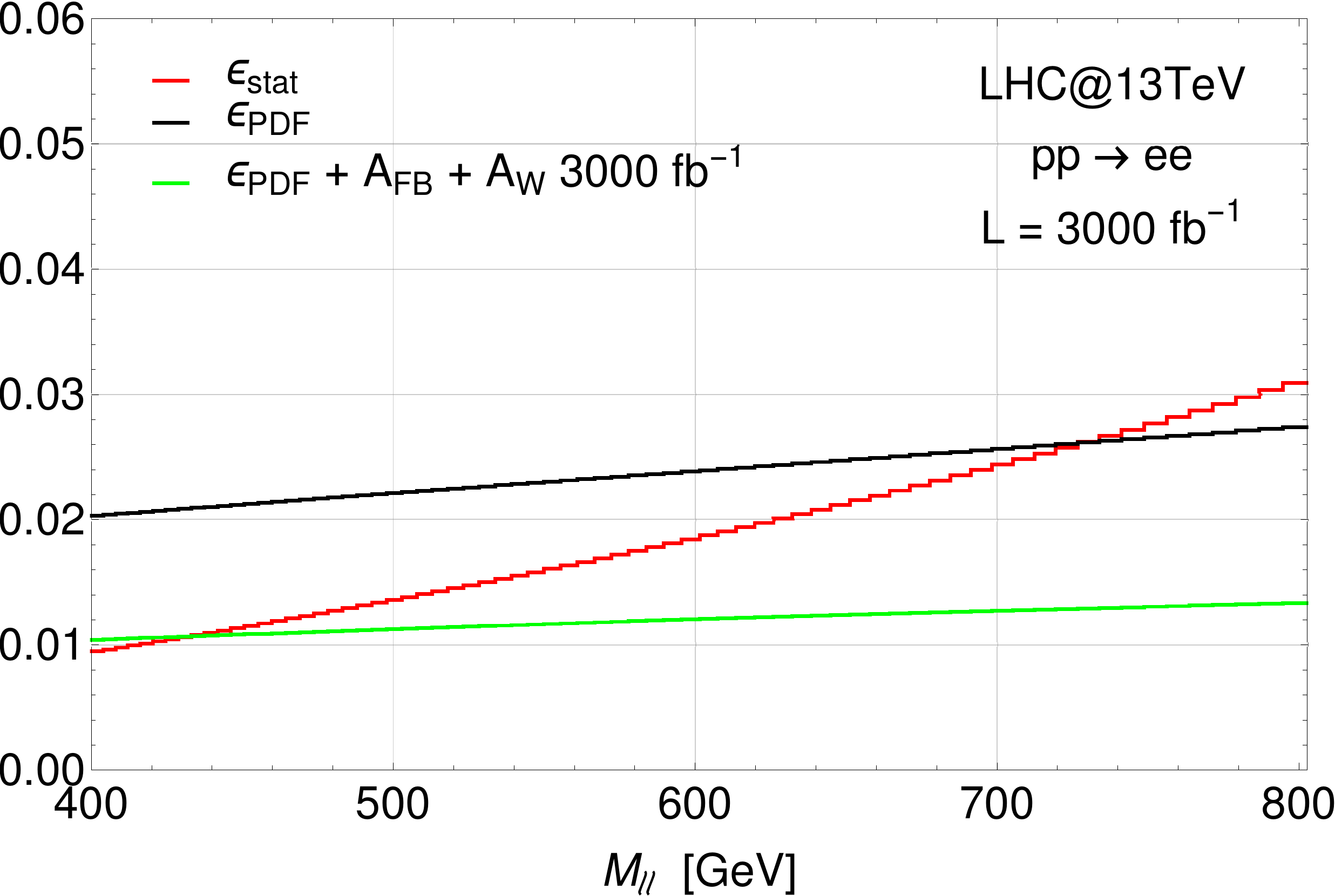}
\includegraphics[width=0.49\textwidth]{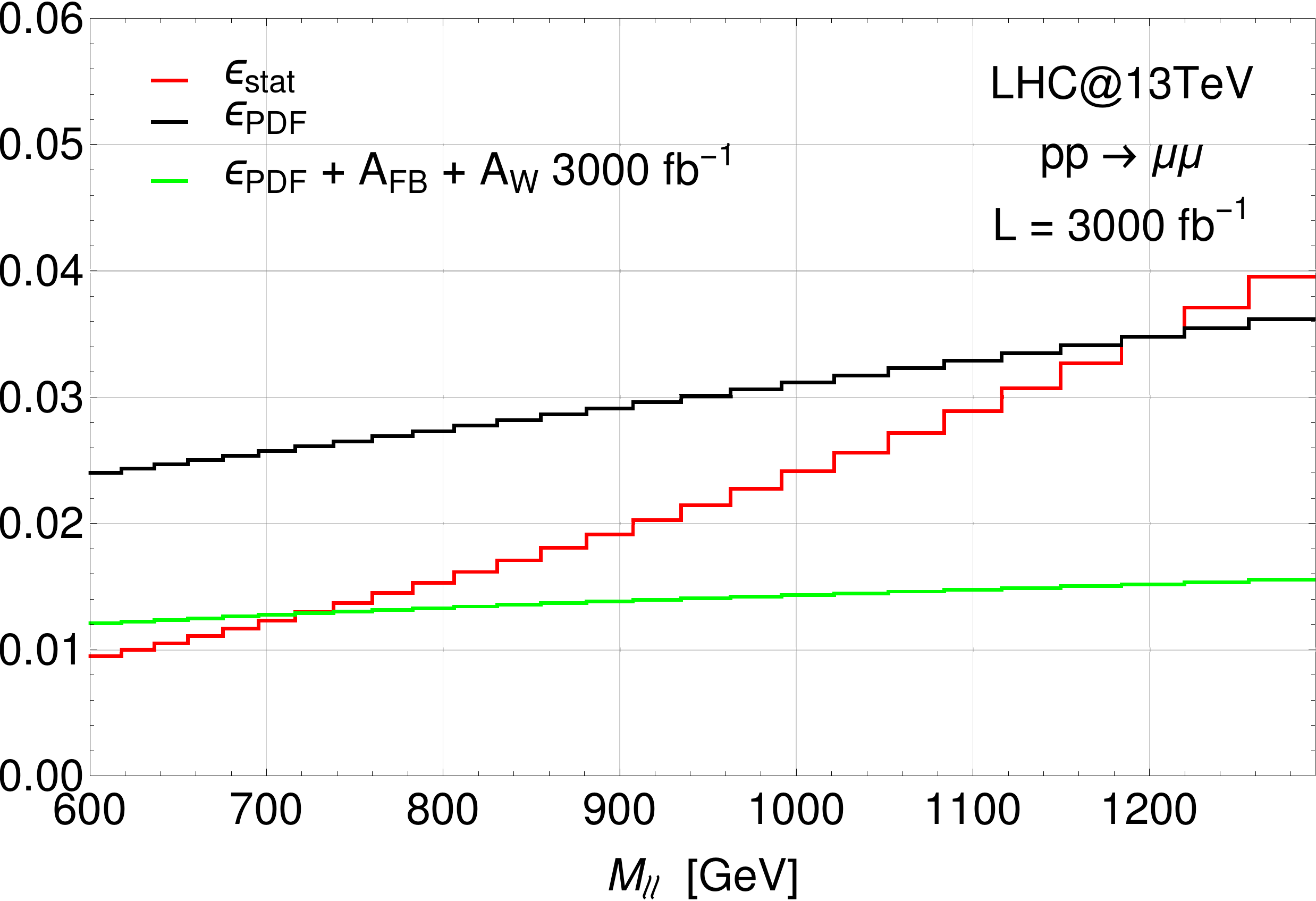}
\end{center}
\caption{Statistical (red) and PDF uncertainties before (black) and after (green) the profiling in the di-electron (left) and di-muon (right) channels. Experimental resolutions determine the choice of the bin sizes and the experimental efficiency of each channel in the barrel-barrel phase space is also included, as declared by CMS~\cite{Sirunyan:2018exx}.}
\label{fig:neutral_rel_error_dilepton}
\end{figure}

For this purpose, in Fig.~\ref{fig:neutral_rel_error_dilepton}, we identify the invariant mass region where statistical and PDF uncertainties are comparable. As intimated, here, we have assumed an integrated luminosity of 3000 fb$^{-1}$ both in the estimation of the statistical errors on the spectrum (red curves) and in the statistical accuracy of the $A_{FB}$ and $A_W$ pseudodata included in the profiling of the PDFs (green curves).
The two are compared with the original CT18NNLO PDF uncertainty (black curves) in the the di-electron (left) and di-muon (right) final states.
The choice of the bin sizes in the two plots reflects the different resolutions of the two channels (about 1\% for electrons and 3\% for the muons) and the estimation of the statistical uncertainties also includes their different efficiencies (about 69\% for electrons and 93\% for muons in the barrel-barrel phase space), e.g., as reported by CMS~\cite{Sirunyan:2018exx}.

\begin{figure}
\begin{center}
\includegraphics[width=0.49\textwidth]{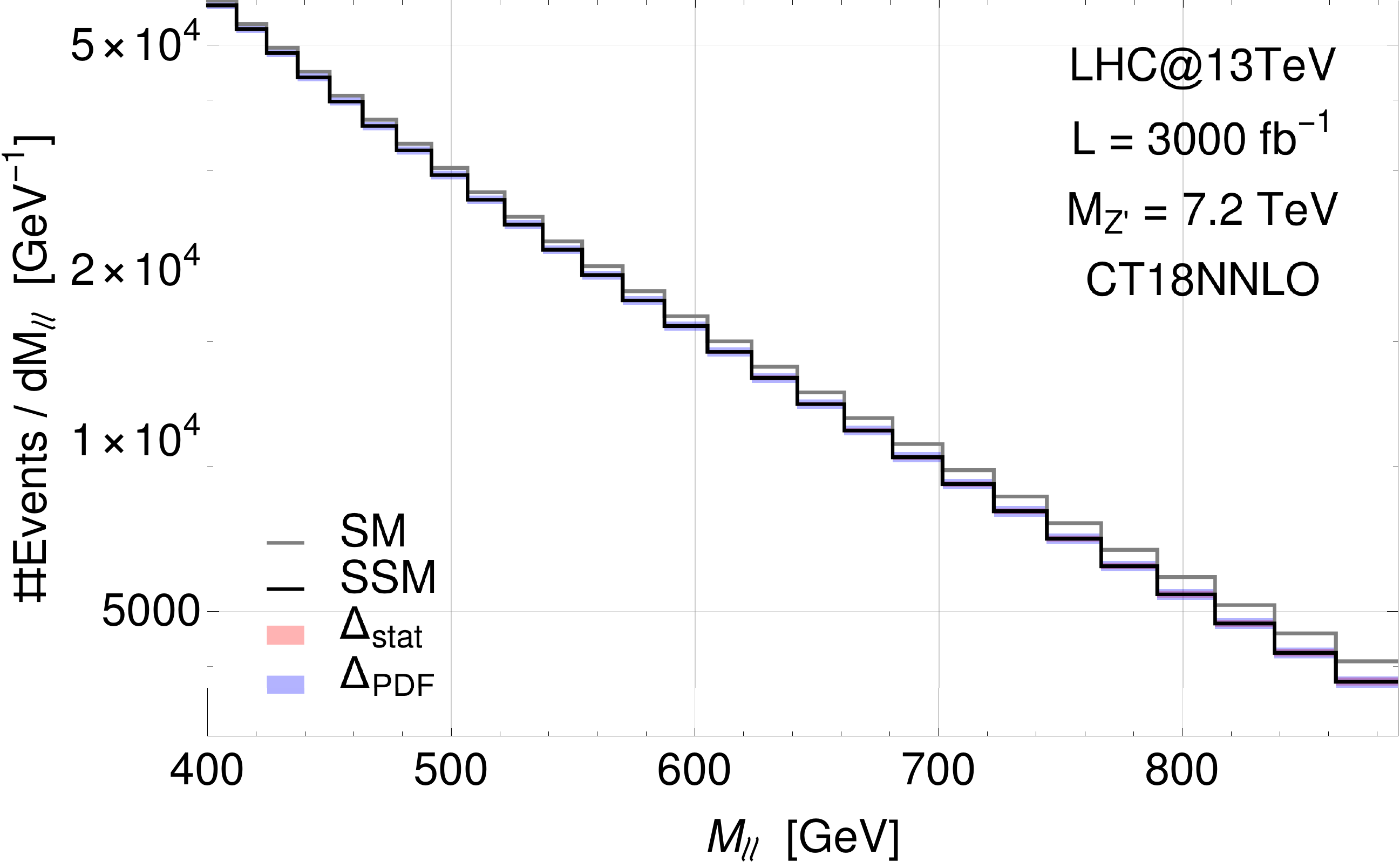}
\includegraphics[width=0.49\textwidth]{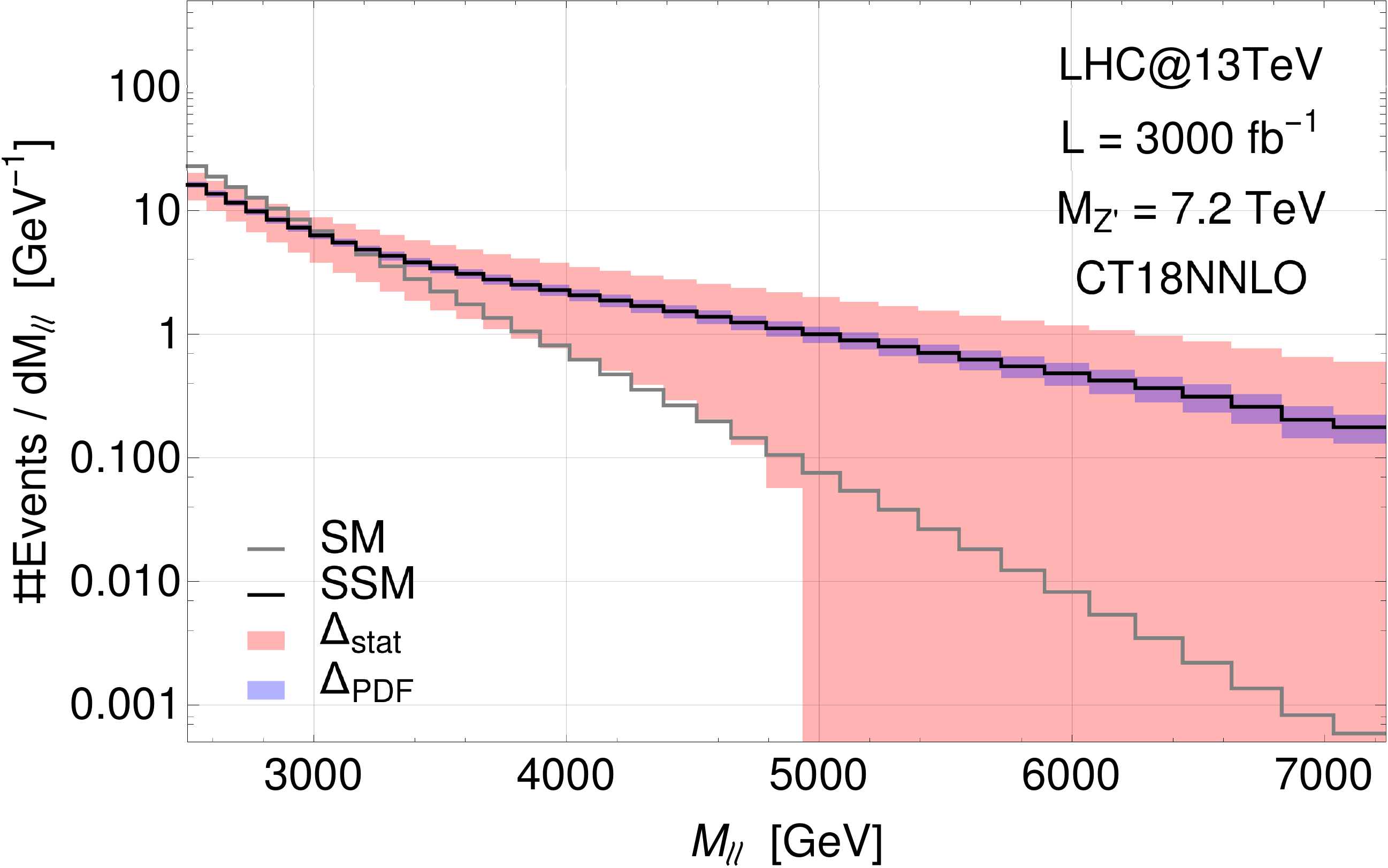}
\end{center}
\caption{Invariant mass distribution of the number of events for the enhanced SSM $Z^\prime$ benchmark with a mass of 7.2 TeV. The PDF uncertainty (blue shade) represents the original CT18NNLO error while the statistical error (red shade) corresponds to 3000 fb$^{-1}$ of integrated luminosity. NNLO QCD corrections have been applied through a $K$-factor. No detector efficiencies are included.}
\label{fig:Z_prime_enhanced}
\end{figure}

A potential BSM signal in this invariant mass region would then be strongly affected by PDF uncertainties, which the inclusion of $A_{FB}$ and $A_W$ data can ameliorate 
significantly by about a factor 2.
To address this effect, we consider the enhanced SSM model introduced in the previous subsection and study its phenomenology in the neutral dilepton channel, where a wide $Z^\prime$ signal arises.
The invariant mass profile, in two specific intervals, of such $Z^\prime$ realisation is visible in Fig.~\ref{fig:Z_prime_enhanced}.

\begin{figure}
\begin{center}
\includegraphics[width=0.49\textwidth]{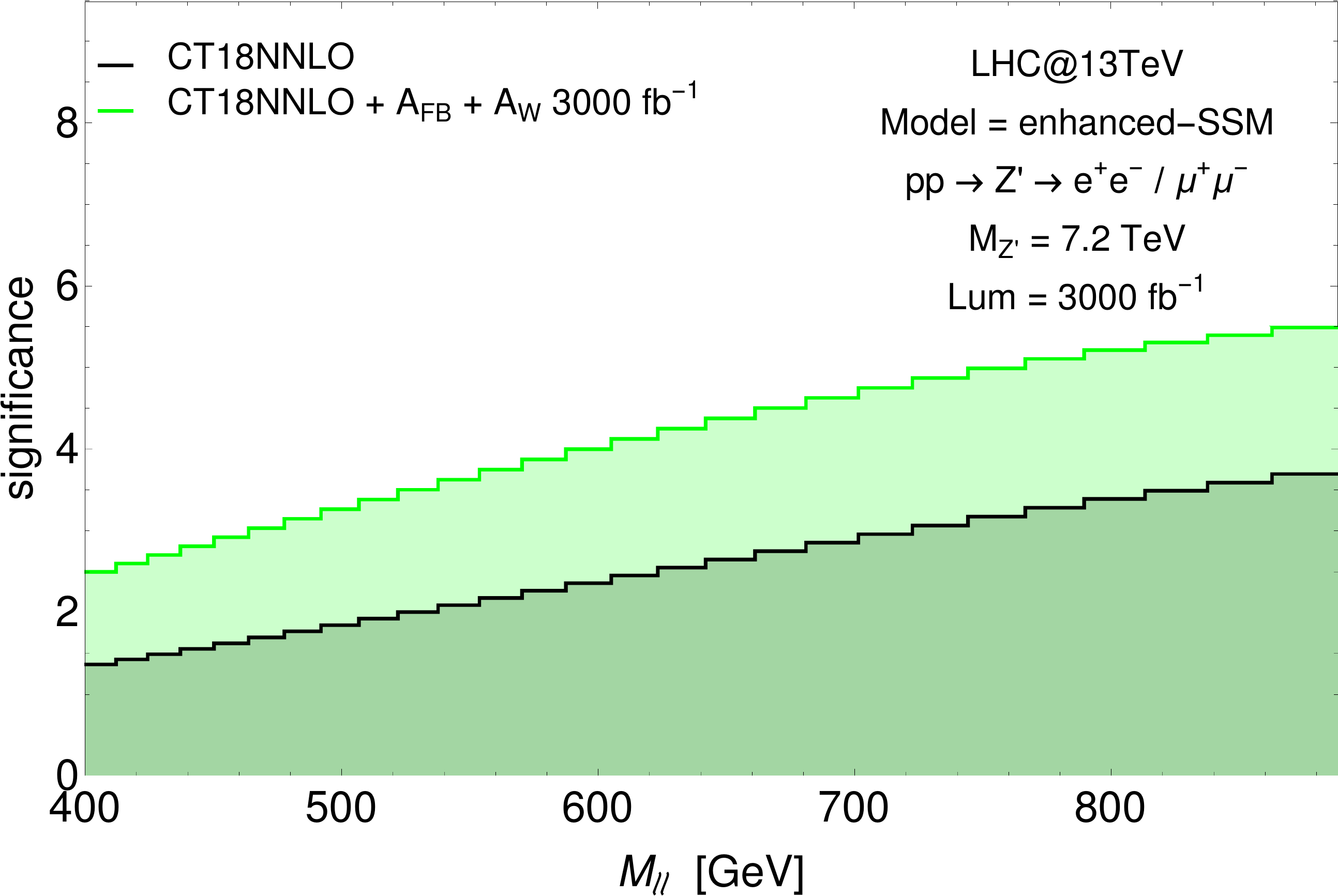}
\includegraphics[width=0.49\textwidth]{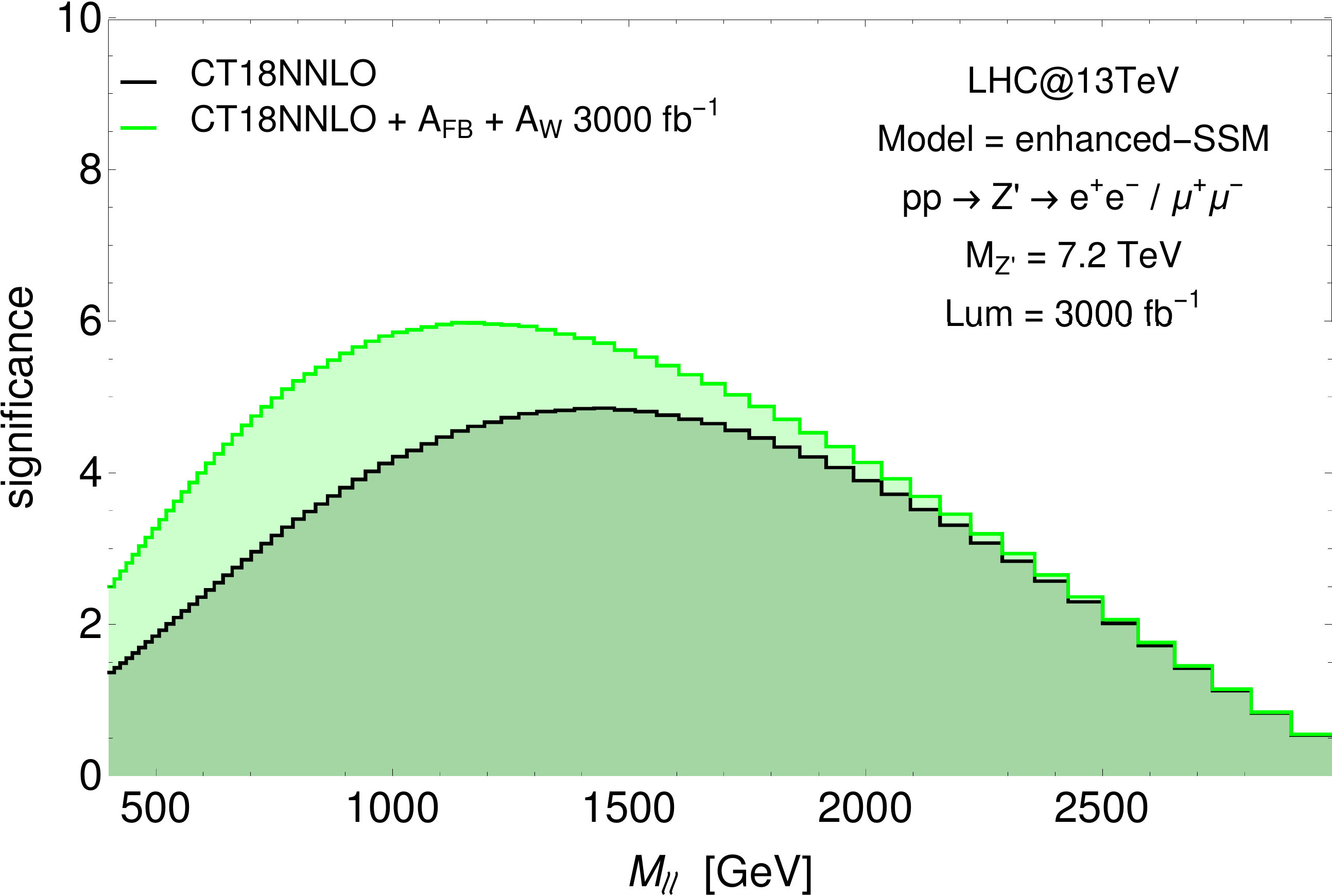}
\end{center}
\caption{Significance of the enhanced SSM $Z^\prime$ signal with an integrated luminosity of 3000 fb$^{-1}$, including the PDF error from original CT18NNLO PDF set (black) and after the profiling (green) with $A_{FB}$ and $A_W$ pseudodata with the same integrated luminosity accuracy. Efficiencies of both lepton channels are included.}
\label{fig:Z_prime_enhanced_sig}
\end{figure}

Assuming an integrated luminosity of 3000 fb$^{-1}$, the overall significance of the broad peak of a resonance of this kind would be about 3.2$\sigma$s when including the PDF error, summed in quadrature with the statistical uncertainty, thus at the edge of the sensitivity.
The depletion of events due to interference effects that appears below the peak, however, would be statistically significant before the actual observation of the peak itself.
Particularly, in the invariant mass region where statistical and PDF uncertainties are comparable, the significance of the depletion of events (assumed here at the same level as an excess of events) would greatly benefit from the foreseen improvement on the PDF error.
This is visible in Fig.~\ref{fig:Z_prime_enhanced_sig}, where the significance of the depletion of events is shown in two invariant mass intervals, for the original CT18NNLO PDF set and for the profiled PDF set using $A_{FB}$ and $A_W$ pseudodata with 3000 fb$^{-1}$ statistical uncertainty.

\subsection{Effects on $W^\prime$ searches}

\begin{figure}
\begin{center}
\includegraphics[width=0.49\textwidth]{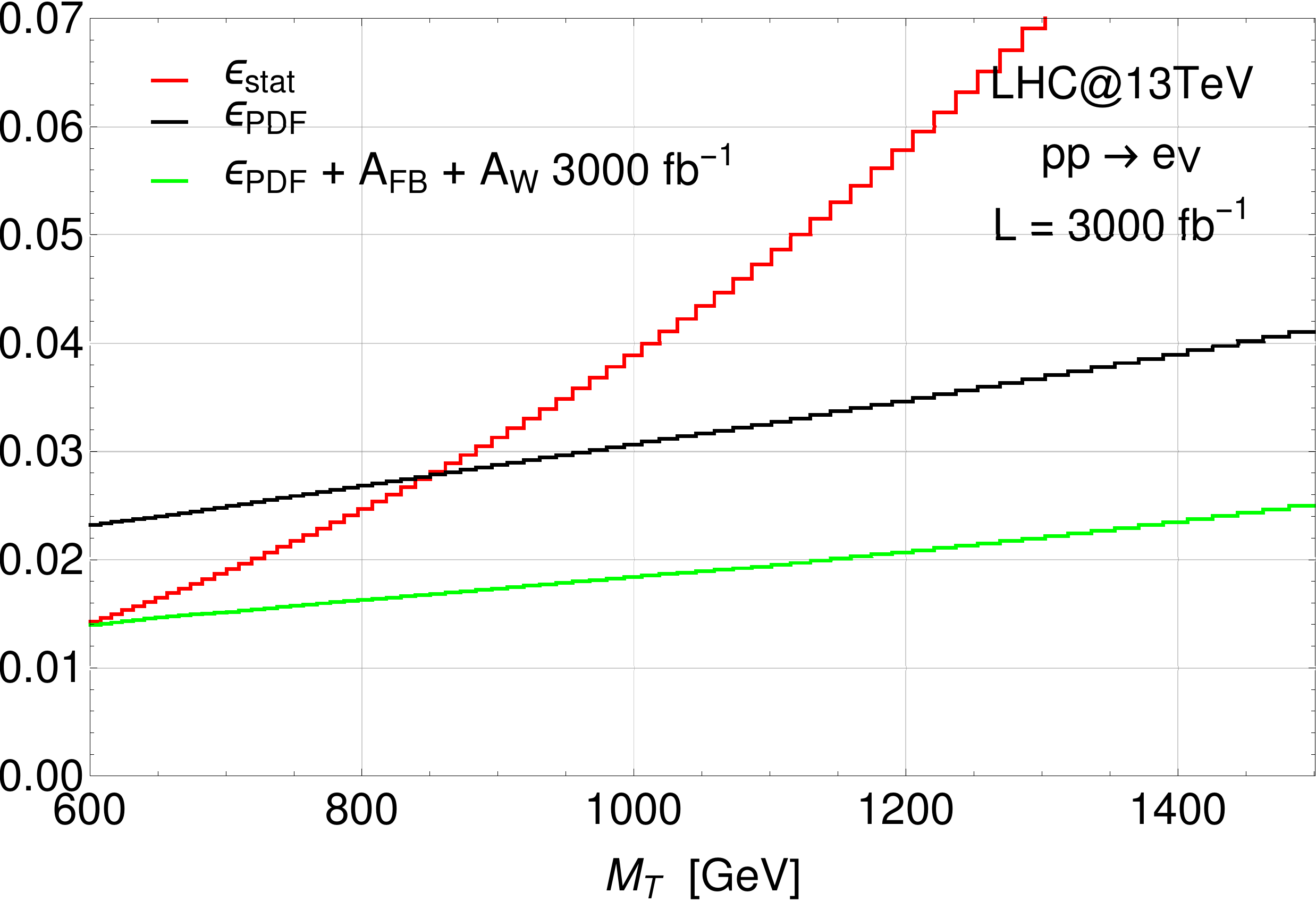}
\includegraphics[width=0.49\textwidth]{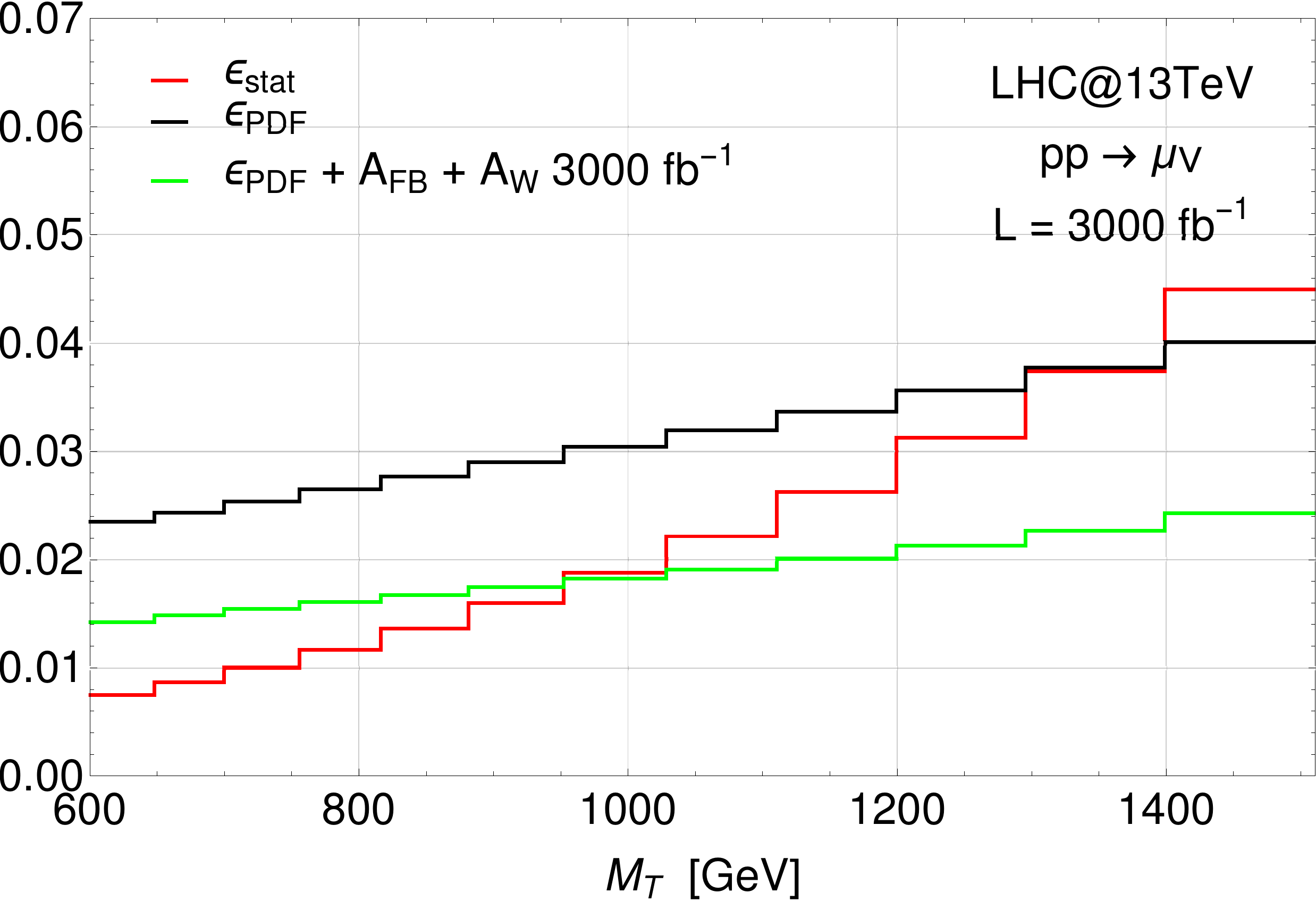}
\end{center}
\caption{Statistical (red) and PDF uncertainties before (black) and after (green) the profiling in the electron (left) and muon (right) plus missing transverse energy channels. Experimental resolutions determine the choice of the bin sizes and the experimental efficiency of each channel is also included, as declared by ATLAS~\cite{Aad:2019wvl}.}
\label{fig:charged_rel_error_dilepton}
\end{figure}

Similar conclusions can be drawn in the charged channel.
Fig.~\ref{fig:charged_rel_error_dilepton} shows the comparison between statistical and PDF uncertainties in the transverse mass distribution in the lepton plus missing transverse energy channel.
Again, an integrated luminosity of 3000 fb$^{-1}$ is assumed in the statistical uncertainty of the spectrum (red curves) and in the accuracy of the $A_{FB}$ and $A_W$ pseudodata included in the profiling of the PDFs (green curves).
The left (right) plot shows the comparison of the uncertainties in the electron (muon) plus missing transverse energy channel, for which a resolution of about 1.3\% (8\%) and an efficiency of about 64\% (44\%) have been assumed, in order to resemble, e.g., the ATLAS experiment features declared in Ref.~\cite{Aad:2019wvl}.

\begin{figure}
\begin{center}
\includegraphics[width=0.49\textwidth]{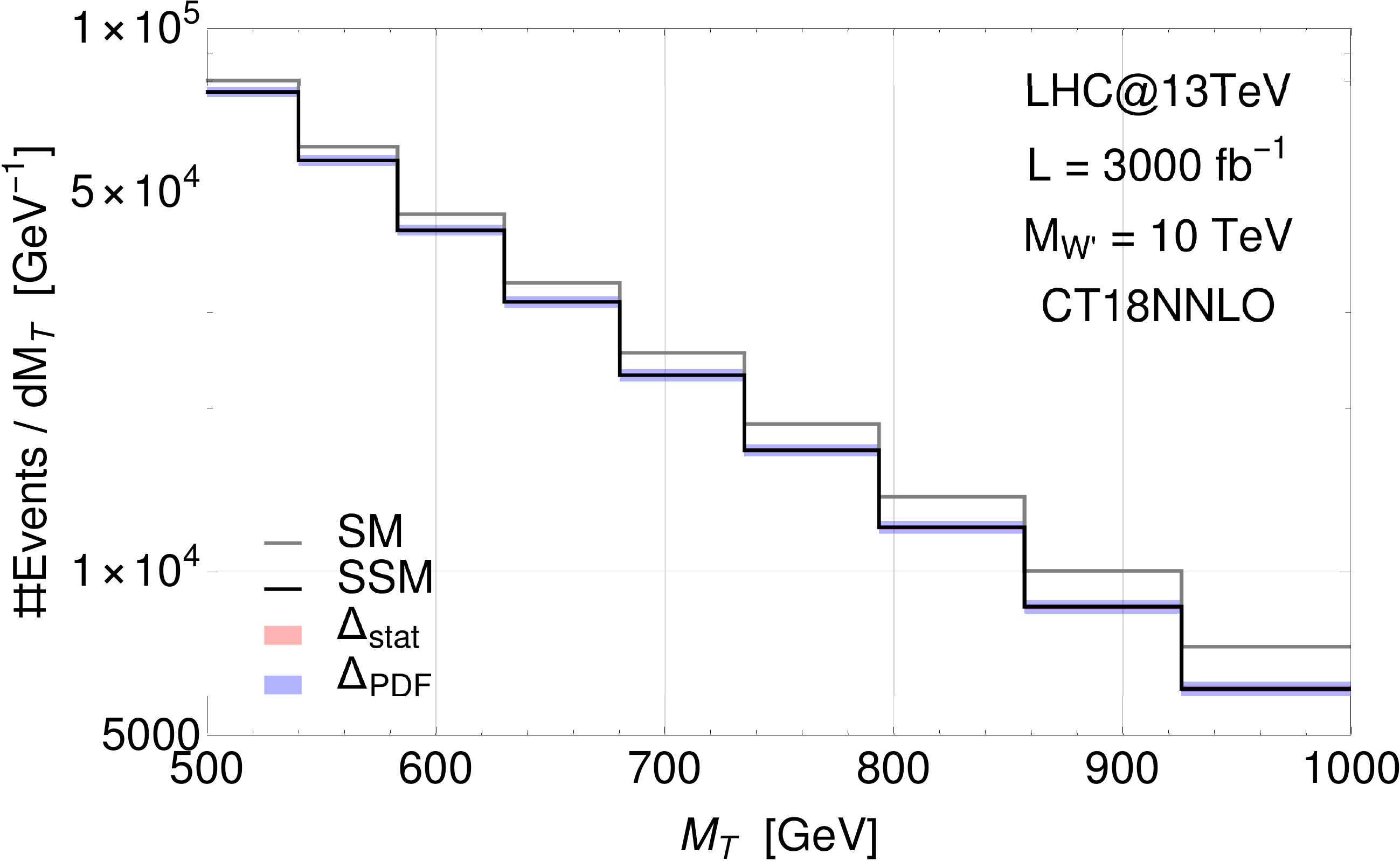}
\includegraphics[width=0.49\textwidth]{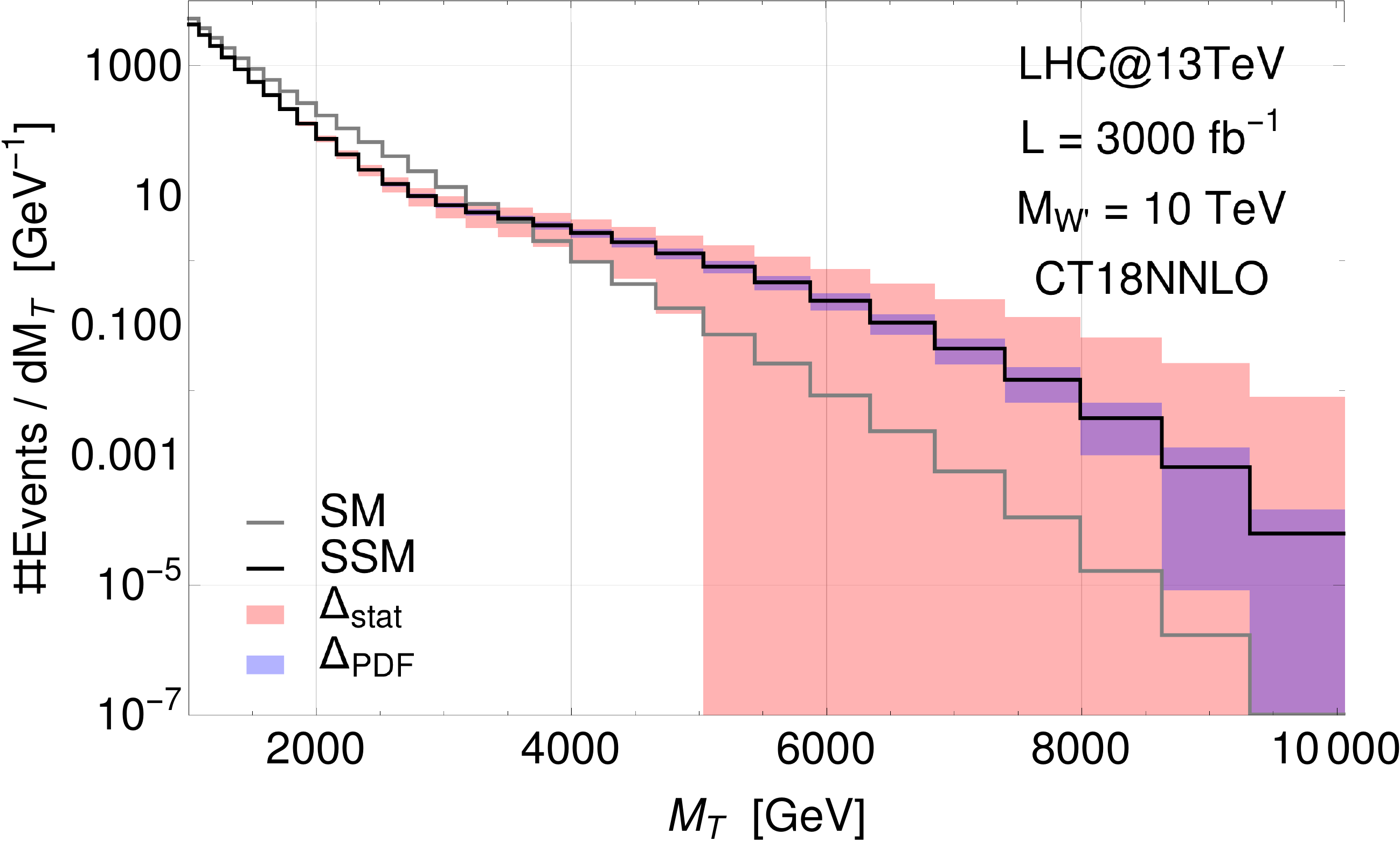}
\end{center}
\caption{Transverse mass distribution of the number of events for the enhanced SSM $W^\prime$ benchmark with a mass of 10 TeV. The PDF uncertainty (blue shade) represents the original CT18NNLO error, while the statistical error (red shade) corresponds to 3000 fb$^{-1}$ of integrated luminosity. No detector efficiencies are included.}
\label{fig:W_prime_enhanced}
\end{figure}

The inclusion of $A_{FB}$ and $A_W$ data can ameliorate the PDF uncertainty by about a factor 2 and this in turn can significantly improve the experimental sensitivity 
to $W^\prime$ states in BSM searches. We study the impact of this effect on the phenomenology of a wide $W^\prime$ resonance in the enhanced SSM scenario already introduced, for which the transverse mass distribution is visible in Fig.~\ref{fig:W_prime_enhanced}.

\begin{figure}
\begin{center}
\includegraphics[width=0.49\textwidth]{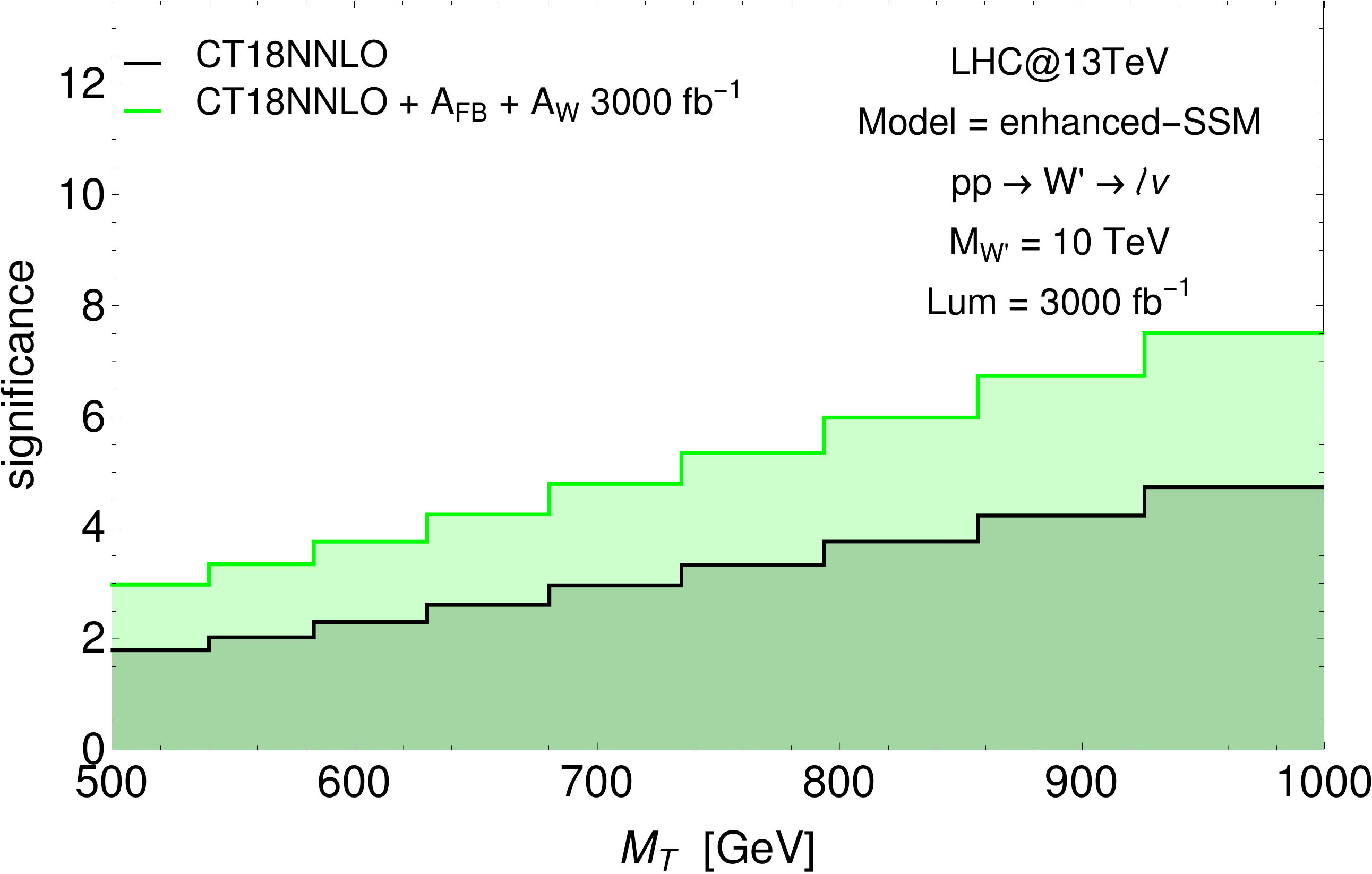}
\includegraphics[width=0.49\textwidth]{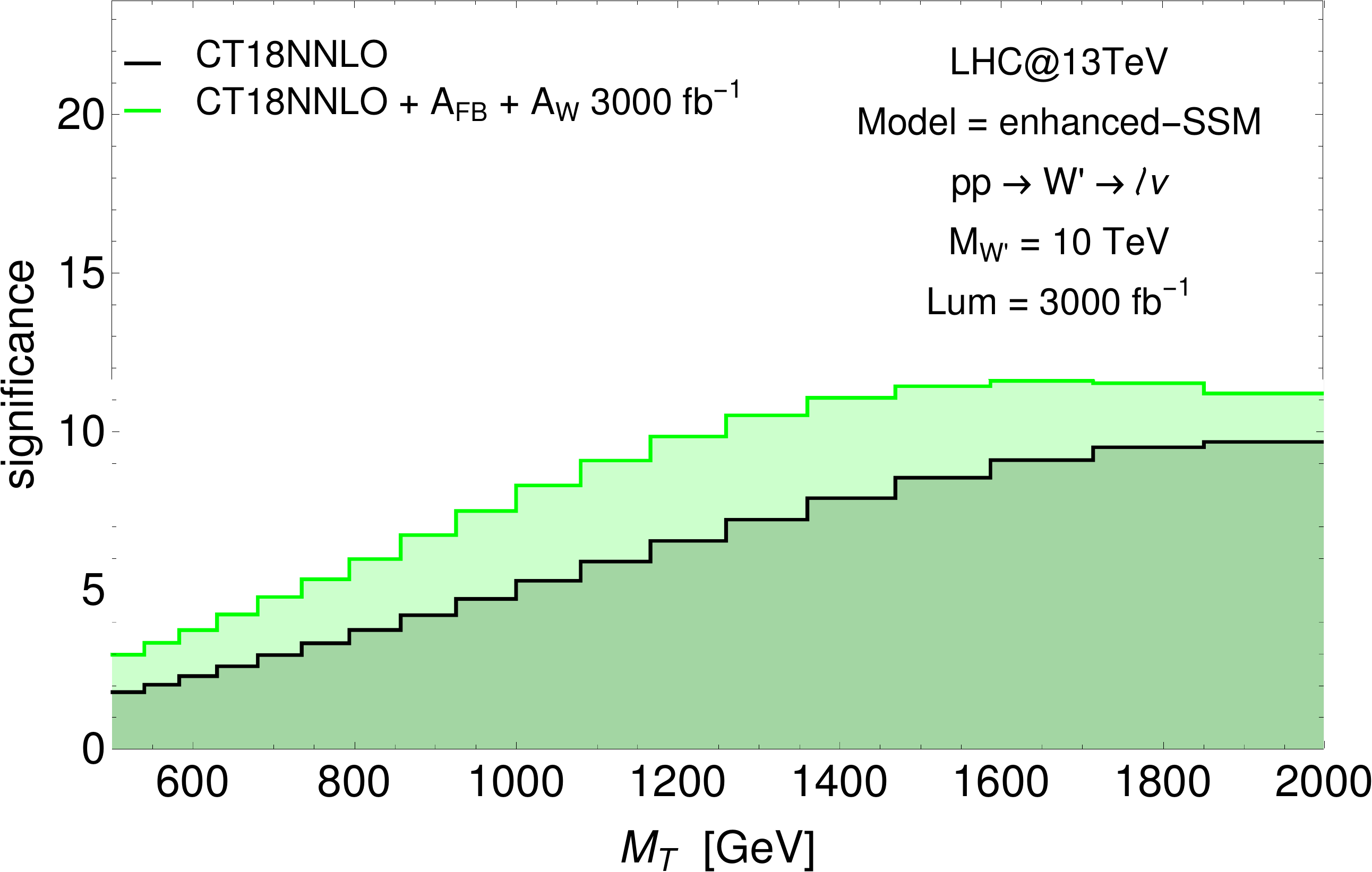}
\end{center}
\caption{Significance of the enhanced SSM $W^\prime$ signal with and integrated luminosity of 3000 fb$^{-1}$, including the PDF error from original CT18NNLO PDF set (black) and after the profiling (green) with $A_{FB}$ and $A_W$ pseudodata with the same integrated luminosity accuracy. Efficiencies of both lepton channels are included.}
\label{fig:W_prime_enhanced_sig}
\end{figure}

With an integrated luminosity of 3000 fb$^{-1}$, the broad peak of the chosen benchmark would be below the HL-LHC sensitivity.
However, the depletion of events due to interference effects would be visible at an early stage in the low transverse mass tail.
In particular, in the transverse mass region where statistical and PDF uncertainties are comparable, the inclusion of $A_{FB}$ and $A_W$ pseudodata with 3000 fb$^{-1}$ would significantly increase the significance of such a signal (again assumed at the same level as an excess of events), as visible in Fig.~\ref{fig:W_prime_enhanced_sig}. 

\section{Conclusions} 
\label{sec:conc} 

The main limitation in precision measurements of EW parameters at the various runs of the LHC, including the HL-LHC option, comes from non-perturbative PDF uncertainties. These are dealt with in many current analyses by in-situ reweighting and profiling techniques. In this work, we have explored the potential of approaches using the combination of asymmetry measurements in the neutral-current and charged-current DY channels to improve our knowledge of PDFs and reduce the corresponding uncertainties. 

While presenting results for a subset of PDFs, we have however verified that the effects described here are of relevance for any PDF set.

Extending the study of Ref.~\cite{Abdolmaleki:2019qmq} using $A_{FB}$, here, we have investigated the impact of combining $A_{FB}$ with the lepton-charge asymmetry $A_W$ and shown, by a quantitative analysis using {\tt{xFitter}}, the complementary constraints provided by the two asymmetries. 
We have first validated our {\tt{xFitter}} implementation against ATLAS data at 8 TeV extracting $A_W$, finding good $\chi^2$ values for all the PDF sets examined, and we have then used $A_{FB}$ and $A_W$ pseudodata at 13 TeV taken near the vector boson ($Z$ and $W$) masses to determine the reduction of PDF uncertainties by considering two luminosity scenarios, of 300 fb$^{-1}$ (appropriate for Run 3 of the LHC) and 3000 fb$^{-1}$ (appropriate for the HL-LHC). 

By eigenvector rotation we have illustrated the role of combining $A_{FB}$ and $A_W$ in order to place constraints on linearly independent combinations of $u$ and $d$ valence quark densities, $(2/3) u_V + (1/3) d_V$ and $u_V - d_V$, respectively, to which the two asymmetries are sensitive at lowest order. Furthermore, we have shown that including $A_{FB}$ and $A_W$ in PDF fits reduces the error on the ratio $\bar d/\bar u$ at very large $x$ values, which has been confirmed by several measurements to be much larger than 1 (thereby hinting at a significant flavour asymmetry in the proton sea of antiquarks), to the extent that experimental data are no longer within the error bands predicted by the PDFs.

We have then found that the combined effect of $A_{FB}$ and $A_W$ leads to a 30\% improvement in the PDF uncertainties on the transverse mass and lepton $p_T$ distributions in lepton-neutrino final states over a broad kinematic range measured at the LHC around the vector boson peak. In fact, the effect of $A_W$ alone in the charged DY channel is not dissimilar from that of $A_{FB}$ in the neutral DY channel, previously seen in the aforementioned reference, so as to suggest that the combined effect of the two asymmetries will also further benefit studies of EW parameters in dilepton final states, chiefly, of $\sin^2\theta_W$ (this is currently being assessed quantitatively). In fact, we also remark here that the improvement of the PDF uncertainties due to the inclusion of $A_{FB}$ and $A_W$ constraints in the mass regions close to $m_Z$ and $m_W$, mapped in the invariant and transverse masses of the neutral and charged channel final states, respectively, may eventually induce a reduction in the error on the determination of $m_Z$ and $m_W$ at the (HL-)LHC (and hadron colliders in general). While this may have little phenomenological impact in the case of the $Z$ mass (which best determination is still given 
by $e^+e^-$ data), it may be of relevance for the $W$ mass (measured more accurately at hadron colliders yet dominated by PDF uncertainties). However, this is beyond the scope of this work and we leave it for a forthcoming publication.

Finally, we have investigated the impact of pseudodata from the peak region on the description of the high-mass (multi-TeV) region in both the neutral and charged DY channel. We have found that the constraints coming from the $A_{FB}$ and $A_W$ combination improve the relative PDF uncertainty by around 20\% in the invariant and transverse mass spectra, respectively, between 2 TeV and 4 TeV, a region where evidence of, e.g., wide $Z^\prime$ and $W^\prime$ states can first be established. 

\section*{Acknowledgements}

We thank S.~Amoroso, S.~Camarda and A.~Glazov for many useful discussions. FH acknowledges the hospitality and support of DESY, Hamburg and CERN, Theory Division. SM is supported in part through the NExT Institute and acknowledge funding from the STFC Consolidated Grant ST/L000296/1. The work of JF has been supported by STFC under the Consolidated Grant ST/T000988/1. JF and SM acknowledge the use of the IRIDIS High Performance Computing facility, and associated support services, at the University of Southampton, in the completion of this work.

%%%%%%%%%%%%%%%%%%%%%%%%%%%%%%%%%%%%%%%%%%%%%%%%%%%%%%%%%%%%%%%%%%%%%%%%%%%%%%%%

\clearpage
\bibliography{references}

%%%%%%%%%%%%%%%%%%%%%

\end{document}